\documentclass[10pt,journal,twocolumn,romanappendices]{IEEEtran}

\IEEEoverridecommandlockouts

\usepackage{silence}
\usepackage{amsthm}
\usepackage{amssymb}
\usepackage{algorithm,algorithmic,amsmath,bbm,cite,color,comment,graphicx,setspace,subcaption,tabularx,url}
\usepackage[USenglish]{babel}
\usepackage{booktabs,tabularx,multirow,array}
\usepackage[table]{xcolor} 
\usepackage{tabularray}
\usepackage[T1]{fontenc}
\usepackage[utf8]{inputenc}
\usepackage[hidelinks]{hyperref}
\usepackage{mathtools}
\usepackage{xfrac}
\usepackage{microtype}
\usepackage{mathrsfs}
\usepackage{csquotes}
\usepackage{makecell}
\usepackage{multirow}
\usepackage{mleftright}
\usepackage{enumitem}
\setitemize{itemsep=0mm,topsep=0mm,parsep=0pt,partopsep=0pt}

\usepackage{pgfplots}
\pgfplotsset{compat=1.18}
\usepackage{tikz}
\usetikzlibrary{patterns,arrows,backgrounds,calc,intersections,decorations.pathreplacing,positioning,shapes}
\usepgfplotslibrary{fillbetween}

\makeatletter
\newcommand\subparagraph{%
  \@startsection{subparagraph}{5}
  {\parindent}
  {3.25ex \@plus 1ex \@minus .2ex}
  {-1em}
  {\normalfont\normalsize\bfseries}}
\makeatother
\usepackage{titlesec}
\let\subparagraph\relax

\usepackage{titlesec}
\let\subparagraph\relax
\titlespacing{\section}{0pt}{5pt plus 2pt minus 1pt}{3pt plus 1pt minus 0pt}
\titlespacing{\subsection}{0pt}{4pt plus 2pt minus 1pt}{2pt plus 1pt minus 0pt}
\usepackage[figurename=Fig.,font=small]{caption}

{\newtheorem{theorem}{Theorem}}
{}
{\newtheorem{lemma}{Lemma}}
{\newtheorem{proposition}{Proposition}}
{\newtheorem{remark}{Remark}}
{}
{\newtheorem{corollary}{Corollary}}

\newcommand\black[1]{\textcolor{black}{#1}}
\def\real{\ensuremath{\mathbb{R}}}
\def\complex{\ensuremath{\mathbb{C}}}

\def\setS{\ensuremath{\mathcal{S}}}
\def\setC{\ensuremath{\mathcal{C}}}
\def\setW{\ensuremath{\mathcal{W}}}
\def\setA{\ensuremath{\mathcal{A}}}

\newcommand\LT[1]{\mathcal{L}\mleft\{#1\mright\}}

\def\setN{\ensuremath{\mathcal{N}}}
\def\setU{\ensuremath{\mathcal{U}}}
\def\setE{\ensuremath{\mathcal{E}}}

\def\n{\ensuremath{\mathbf{n}}}

\def\r{\ensuremath{\mathbf{r}}}

\def\y{\ensuremath{\mathbf{y}}}

\def\T{\ensuremath{\mathbf{T}}}

\def\h{\ensuremath{\mathbf{h}}}

\def\I{\ensuremath{\mathbf{I}}}

\def\1{\ensuremath{\mathbf{1}}}
\def\0{\ensuremath{\mathbf{0}}}

\def\herm{\ensuremath{{\rm H}}}
\def\tran{\ensuremath{{\rm T}}}

\newcommand\norm[1]{\mleft\lVert#1\mright\rVert}
\newcommand\abs[1]{\lvert#1\rvert}

\def\max{\ensuremath{\mathrm{max}}}
\def\min{\ensuremath{\mathrm{min}}}

\newcommand{\re}{{\mathfrak{R}}}
\newcommand{\im}{{\mathfrak{I}}}

\newcommand\Exp[1]{\mathbb{E}\mleft\{#1\mright\}}

\newcommand\Expp[2]{\mathbb{E}_{#1}\mleft\{#2\mright\}}

\newcommand\Prob[1]{\mathbb{P}\mleft\{#1\mright\}}
\newcommand{\sgn}{\mathrm{sgn}}
\newcommand\qfunc[1]{Q\mleft(#1\mright)}
\newcommand\qfuncsq[1]{Q^2\mleft(#1\mright)}
\newcommand\gammaf[1]{\Gamma\mleft(#1\mright)}

\newcommand\ssin[1]{\sin#1}

\newcommand\qnbitp[2]{\ensuremath{\mathscr{Q}_{#1}\mleft(#2\mright)}}

\def\widetildetheta{\widetilde{\rule{0pt}{1\fontcharht\font`B}\smash{\theta}}}

\def\SEPG{\mathscr{P}}
\def\SEP{\SEPG_{\scriptscriptstyle \textup {QPSK}}}
\def\SEPBPSK{\SEPG_{\scriptscriptstyle \textup {BPSK}}}
\def\SEPSISO{\SEPG_{\scriptscriptstyle \textup {QPSK}}^{\scriptscriptstyle \textup {SISO}}}
\def\mrc{{\scriptscriptstyle \textup {MRC}}}
\def\sc{{\scriptscriptstyle \textup {SC}}}
\def\uq{{\scriptscriptstyle \textup {UQ}}}

\def\miso{{\scriptscriptstyle \textup {MISO}}}
\def\summa{{\scriptscriptstyle \textup {sum}}}
\def\icsi{{\scriptscriptstyle \textup {LCSI}}}
\def\md{{\scriptscriptstyle \textup {MD}}}

\def\nbit{{\scriptscriptstyle n\ge3}}
\def\twobit{{\scriptscriptstyle n=2}}
\def\equalcase{{\scriptscriptstyle M = 2^n}}
\def\inequalcase{{\scriptscriptstyle M < 2^n}}
\def\efcase{{\scriptscriptstyle M > 2^n}}
\def\onebit{{\scriptscriptstyle n=1}}
\def\dg{{\scriptscriptstyle \textup {d}}}
\def\cg{{\scriptscriptstyle \textup {c}}}
\def\lb{{\scriptscriptstyle \textup {L}}}
\def\ub{{\scriptscriptstyle \textup {H}}}

\newcommand\br[1]{\mleft(#1\mright)}

\newcommand{\Herm}{^{\mathrm{H}}}
\newcommand{\Trans}{^\mathrm{T}}

\def\Sigmab{\ensuremath{\boldsymbol{\Sigma}}}

\DeclarePairedDelimiter\ceil{\lceil}{\rceil}
\DeclarePairedDelimiter\floor{\lfloor}{\rfloor}
\DeclareMathOperator*{\argmax}{argmax}
\DeclareMathOperator*{\argmin}{argmin}

\def\tfo{{_2F_1}}

\newcommand\smallO[1]{
    \mathchoice
    {
      {\scriptstyle\mathcal{O}}{\mleft(#1\mright)}
    }
    {
      {\scriptstyle\mathcal{O}}{\mleft(#1\mright)}
    }
    {
      {\scriptscriptstyle\mathcal{O}}{\mleft(#1\mright)}
    }
    {
      \scalebox{0.8}{$\scriptscriptstyle\mathcal{O}$}{\mleft(#1\mright)}
    }
}

\newcommand*{\ncr}[2]{
\mleft(
\mathchoice
    {
      \genfrac{}{}{0pt}{0}{\smash{#1}}{\smash[b]{#2}}
    }
    {
      \genfrac{}{}{0pt}{1}{\smash{#1}}{\smash[b]{#2}}
    }
    {
      \genfrac{}{}{0pt}{2}{\smash{#1}}{\smash[b]{#2}}
    }
    {
      \genfrac{}{}{0pt}{3}{\smash{#1}}{\smash[b]{#2}}
    }
    \mright)
}

\newcommand*{\ncrr}[2]{
\mleft(
\mathchoice
    {
      \genfrac{}{}{0pt}{0}{{#1}}{{#2}}
    }
    {
      \genfrac{}{}{0pt}{1}{{#1}}{{#2}}
    }
    {
      \genfrac{}{}{0pt}{2}{{#1}}{{#2}}
    }
    {
      \genfrac{}{}{0pt}{3}{{#1}}{{#2}}
    }
    \mright)
}

\colorlet{kalu}{black}

\definecolor{left} {HTML}{dddddd}
\definecolor{tteal} {HTML}{bee2cd}
\definecolor{tteall} {HTML}{71e1af}
\definecolor{creem} {HTML}{fdcdac}
\definecolor{bluu} {HTML}{cbd5ff}
\definecolor{pinnk} {HTML}{f4cae4}
\definecolor{dark} {HTML}{768196}

\title{SEP Analysis of a Low-Resolution SIMO System with $M$-PSK over Fading Channels}
\author{
Amila Ravinath, Minhua Ding, Bikshapathi Gouda, Italo Atzeni, and Antti Tölli
\thanks{The authors are with the Centre for Wireless Communications, University of Oulu, Finland (e-mail: \{amila.ravinath, minhua.ding, bikshapathi.gouda, italo.atzeni, antti.tolli\}@oulu.fi).}
\thanks{This work was presented in part at ASILOMAR 2025~\cite{C}.}
\thanks{This work was supported by the Research Council of Finland (336449
Profi6, 348396 HIGH-6G, 357504 EETCAMD, and 369116 6G~Flagship) and by the
European Commission (101095759 Hexa-X-II).}}

\begin{document}

\maketitle

\begin{abstract}
In this paper, the average symbol error probability (SEP) of a 
phase-quantized single-input multiple-output (SIMO) system with $M$-ary phase-shift keying (PSK) modulation is analyzed  under Rayleigh fading and additive white Gaussian noise.
By leveraging a novel method, we derive exact SEP expressions for a quadrature PSK (QPSK)-modulated, $n$-bit
phase-quantized SIMO system with maximum ratio combining
(SIMO-MRC), along with the corresponding high signal-to-noise ratio (SNR) characterizations in terms of diversity and coding gains. 
For a QPSK-modulated, $2$-bit phase-quantized SIMO system with selection combining,
the diversity and coding gains are further obtained for an arbitrary
number of receive antennas, complementing existing
results. Interestingly, the proposed method also reveals a duality between a SIMO-MRC system and a phase-quantized
multiple-input single-output (MISO) system
with maximum ratio transmission when the
modulation order, phase-quantization resolution, 
antenna configuration, and the channel state information (CSI) conditions are
reciprocal. This duality enables direct inference to obtain the diversity
of a general $M$-PSK-modulated, $n$-bit phase-quantized SIMO-MRC
system, and extends the results to its MISO counterpart. All the above results are obtained assuming perfect CSI at the receiver (CSIR).
Finally, the SEP analysis of a
QPSK-modulated, $2$-bit phase-quantized SIMO system is extended to the limited CSIR case, where the CSI at each receive antenna is represented by only $2$ bits of channel phase
information. In this scenario, the diversity gain
is shown to be further halved in general.\\
\end{abstract}

\section{Introduction}
Multi-antenna systems with large antenna arrays  have been
recognized as a main pillar among physical-layer technologies
to significantly upgrade the spectral efficiency and reliability
of current and future wireless systems~\cite{Rusek-et-al}. 
However, fully digital implementations of such systems inevitably entail a sharp
increase in hardware cost, complexity, and power consumption. Specifically,
the power consumption of analog-to-digital converters
(ADCs) scales exponentially with the number of quantization
bits~\cite{Emil-etal-2017, Lozano_2021}, which undermines the feasibility of employing high-resolution ADCs in large quantities in massive multi-antenna
systems. Consequently, extensive research efforts have focused on low-resolution systems in general and on
$1$-bit quantized systems in particular. The main research themes include, e.g., capacity characterization and
bounds~\cite{J_Mo_R_Heath, Lozano_2021}, channel estimation and data detection~\cite{Y_Li_et_al_BLMMSE, Atzeni_2022,
Risi_Larsson_arxiv_paper, Choi_ML, Rad26},
and symbol error probability (SEP) analysis~\cite{Gayan_2020_Open_J, Gay21}.

When low-resolution quantization is employed at the receiver,
the established results on unquantized multi-antenna systems need
to be revisited, typically requiring entirely new analytical approaches~\cite{Lozano_2021}. 
In~\cite{J_Mo_R_Heath}, the capacities of single-input single-output (SISO) and
multiple-input single-output (MISO) fading channels with $1$-bit
ADCs at the receiver and perfect channel state information
(CSI) at both the transmitter and the receiver were determined,
whereas the capacities of single-input multiple-output (SIMO)
and multiple-input multiple-output fading channels remain only partially characterized.

Parallel to capacity characterization is the analysis of
the average SEP (or simply SEP), a key system reliability
performance index~\cite{Gia03, Ribeiro_Cai_Giannakis_2005, Ber21, Ang_Li_et_al_2020, Wu_Liu_precoding_2024}. Two important parameters, i.e.,
the diversity gain (also known as the diversity order) and
the coding gain, succinctly delineate the 
vanishing SEP at
high signal-to-noise ratio (SNR)~\cite{Gia03, Ribeiro_Cai_Giannakis_2005}. In~\cite{Gayan_2020_Open_J}, the average
SEP for a SISO system with low-resolution
quantization at the receiver and with $M$-ary phase-shift keying
($M$-PSK) modulation at the transmitter was analyzed. The work in~\cite{Wu_Liu_et_al_2023}
examined the diversity gain of a fading
MISO system employing $M$-PSK modulation and low-resolution
digital-to-analog converters (DACs) at the transmitter, further revealing how the number of quantization bits
and the modulation order jointly affect the achievable diversity
gain.

Unlike in MISO systems, the SEP performance of a SIMO system with low-resolution quantized reception has not been
characterized to the same extent. In~\cite{Gay21}, a quadrature phase-shift
keying (QPSK)-modulated, low-resolution phase-quantized
SIMO system with selection combining (SC) (SIMO-SC) was
studied, and the achievable diversity was partially determined
for the case with $1$-bit ADCs. For multi-channel
signal reception, maximum ratio combining (MRC) has been widely adopted for data detection, as in~\cite{Jac17, Y_Li_et_al_BLMMSE, Risi_Larsson_arxiv_paper, Sun21, Atzeni_2022}. However, to the best of our knowledge, an exact SEP expression
of a phase-quantized SIMO system employing MRC (SIMO-MRC)
has not yet been reported. Furthermore, while the coding gain provides an additional means to
further distinguish two systems achieving the same diversity
gain, it has not been 
discussed in existing studies~\cite{Gayan_2020_Open_J,Gay21,Wu_Liu_et_al_2023,Gay17}. Therefore, the SEP analysis
of low-resolution SIMO-MRC systems along with the corresponding high-SNR characterizations still remains an open problem.

Assuming coherent detection with perfect CSI at the receiver
(CSIR), a commonly used approach (hereafter referred to as
the conventional method) for analyzing the SEP for a SIMO-MRC system follows a two-step process~\cite{Gia03, Sim00}: first, given
a particular channel realization, the conditional SEP that accounts for the effect of noise is derived; then, the
conditional SEP is averaged over the fading statistics. However, in the presence of quantization at the receiver, the
signal and noise components are no longer separable after quantization, making it difficult to obtain the conditional
SEP in the first step and, consequently,
to derive an average SEP
expression in the second step. In short, the SEP analysis of a
low-resolution SIMO-MRC system requires a fundamentally different analytical
approach.

In this work, we focus on the SEP analysis of a low-resolution SIMO
system operating over independent and identically distributed (i.i.d.) Rayleigh fading channels
with a large number of receive antennas. While perfect CSIR is
assumed throughout the paper, we also examine the case with limited CSIR, specifically the extreme scenario where the channel estimates are themselves phase-quantized with $2$ bits. As
in~\cite{Wu_Liu_et_al_2023}, a general $M$-PSK modulation is employed at the
transmitter, whereas the receiver is equipped with $n$-bit phase
quantizers~\black{\cite{Gayan_2020_Open_J, Wu_Liu_et_al_2023, Ber22, Ber21b, Sin13}. Such a receiver observes only the phase sector of the received
signal, which facilitates practical implementation \cite{Sin13}}. It is worth noting that a $2$-bit phase quantizer corresponds to a conventional $1$-bit ADC, as it uses one quantization bit for each of the in-phase and quadrature components. The two terms are therefore used
interchangeably when the context is clear.

The main contributions of this paper are summarized
as follows.
\begin{itemize}
\item To overcome the limitations of the conventional two-step SEP
analysis for a quantized system, we devise a new analytical method for a SIMO-MRC system employing $M$-PSK modulation and $n$-bit phase quantization at the
receiver by jointly leveraging the circular symmetry of both the
noise and fading distributions.
\item \black{We establish a duality in terms of average SEP between the considered SIMO-MRC
system and the MISO system employing quantized maximum ratio transmission (MRT)
studied in~\cite{Wu_Liu_et_al_2023} when both use the same $n$-bit phase
quantizers and $M$-PSK modulation. As a result, the diversity gain derived for the MISO-MRT system
in~\cite[Thm.~2]{Wu_Liu_et_al_2023} directly applies to
the corresponding SIMO-MRC system, while the SEP characterizations developed in this work directly apply to the corresponding
MISO setting. Furthermore, for both dual systems, we characterize the associated coding gains up to a scaling factor bounded between $1$ and $2$ whenever a non-zero diversity gain is achieved, and otherwise derive a lower bound on the resulting error floor.}
\item Based on the new analytical method, an exact analytical SEP expression is derived for a QPSK-modulated, $n$-bit phase-quantized SIMO-MRC system, which is further simplified for $n = 2$. Under the same setting, we also derive an approximate closed-form expression of the SEP.  Both
the diversity and coding gains are derived for
any $n \ge 2$. 
\item For a SIMO-SC system  with QPSK modulation and $1$-
bit ADCs (i.e., $2$-bit phase quantization) at the receiver, we 
derive the diversity and coding gains for any 
number of receive antennas, thereby
complementing the results in~\cite{Gay21}. 
\item Lastly, we extend the SEP analysis to the case with limited CSIR. Specifically, when each
receive antenna is provided with only $2$-bit quantized channel estimates, we apply a majority-decision rule among the antenna branches and derive closed-form expressions for the average SEP, diversity gain, and coding gain for a QPSK-modulated, $2$-bit phase-quantized SIMO system.
\end{itemize}

Part of this work was presented in~\cite{C},
which examined the SEP as well as the diversity and coding
gains for a QPSK-modulated, $2$-bit phase-quantized SIMO-MRC
system under i.i.d. Rayleigh fading, albeit without full proofs. The corresponding results for the 
SIMO-SC system were also
presented therein. In this paper, we provide detailed derivations and additional
insights for the preliminary results in~\cite{C}. More importantly, substantial extensions
beyond~\cite{C} include the new
analytical method for an $M$-PSK-modulated, $n$-bit phase-quantized SIMO-MRC
system, the SEP analysis for a QPSK-modulated, $n$-bit phase-quantized SIMO-MRC
system, the duality between a MISO-MRT and a SIMO-MRC system when both use the
same $n$-bit phase quantizers and $M$-PSK modulation, as well as the SEP analysis of a
QPSK-modulated, $2$-bit phase-quantized SIMO system when the CSI at each receive antenna is
limited to $2$ bits of phase information.
\black{Furthermore, a follow-up work~\cite{C2} extends the developed framework to $M$-PSK modulated, $n$-bit quantized SIMO system under correlated Rayleigh fading.}

The rest of the paper is organized as follows. Section~\ref{sec: sys}
introduces the system model and problem statement.
Our new approach to the SEP analysis of a low-resolution phase-quantized SIMO-MRC system is unveiled in Section~\ref{sec: main}.
\black{The MISO-SIMO duality and its implications
are analyzed in Section~\ref{sec: duality}}.
Section~\ref{sec: qpsk} focuses on the cases with QPSK modulation at
the transmitter. Specifically, exact analytical expressions for
the average SEP are derived together with the corresponding diversity and
coding gains when $n$-bit phase quantization is
used at the receiver. This section also includes an
approximate closed-form expression for a SIMO-MRC system
as well as the diversity and coding gains of a SIMO-SC system
when $2$-bit phase-quantizers are used at the receiver. The analysis is extended to the limited CSIR case in Section~\ref{sec: qcsi}, where  the average SEP for a QPSK-modulated, $2$-bit phase-quantized SIMO system with 2-bit quantized channel estimates is derived. Lastly, Section~\ref{sec: concl}
concludes the paper.

\emph{Notation.} Boldface lowercase and uppercase letters represent vectors and matrices, respectively. $\0$ and $\1$ are 
the all-zero vector and all-one vector, respectively and $\I_n$ represents the $n\times n$ identity matrix. $\|\mathbf{a}\|_2$ is the
Euclidean norm of $\mathbf{a}$ and $\|\mathbf{a}\|_1$ its $\ell_1$-norm. $\mathbf{a}^\tran$,
$\mathbf{a}^\herm$, and $\mathbf{a}^*$ represent the transpose, Hermitian transpose, and
element-wise conjugate of $\mathbf{a}$, respectively. $a_n$ denotes
the $n$th entry of the vector $\mathbf{a}$, $|a_n|$ the modulus of $a_n$, and
$\arg(a_n)$ the argument of $a_n$ (defined in $(-\pi, \pi]$). The  imaginary unit is denoted by $j=\sqrt{-1}$.  $\re{(\cdot)}$ and
$\im{(\cdot)}$ denote the real and imaginary parts, respectively. $\ceil*{\cdot}$ and $\floor*{\cdot}$ denote the ceiling and floor operators, respectively. $\emptyset$ denotes the empty set. $\real$ and $\complex$
denote the sets of real and complex values, respectively. The $k$-fold Cartesian
product of a set $\setA$ is denoted by $\setA^k$. $[n]$ denotes the set $\{1, \ldots,
n\}$. $n!$ denotes the factorial of $n$, $\ncr{n}{k} \triangleq
\frac{n!}{(n-k)!k!}$ the binomial coefficient, and $\ncr{\setA}{\ge k}$ the set
of subsets of $\setA$ with size $k$ or more. $\Exp{\cdot}$ and $\Prob{\cdot}$ denote expectation and
probability operators, respectively.  $\setC\setN(\0, \Sigmab)$
represents the zero-mean circularly symmetric complex Gaussian
distribution with covariance matrix $\Sigmab$.  $\setU(a, b)$ denotes the uniform distribution over the
interval $(a, b)$ and $\exp(\lambda)$ the exponential distribution with mean
$\lambda$. 
$\sgn(\cdot)$ stands for the
signum function. 
  $\qfunc{x} \triangleq
\frac1{\sqrt {2\pi}}\int_x^{\infty}e^{-\frac{t^2}2}{dt}$ represents the $Q$-function,
$\Gamma(z) \triangleq \int_0^\infty t^{z-1}e^{-t}\, {dt}$, $\re\{z\}>0$ the
gamma function, $\tfo$ the Gauss hypergeometric function, and $J_0$ the Bessel function of the first kind
and zeroth order. The probability density function (pdf) of a random variable $Z$ is denoted by $f_Z(z)$. $\LT{f(t)}(s) \triangleq \int_{0}^\infty f(t)e^{-st}\ dt$
denotes the Laplace transform of a function $f(t)$. $\varphi_X(\omega) \triangleq \int_{-\infty}^\infty f_X(x)e^{j\omega x}\, dx$ is the characteristic function of the random variable $X$ which has the pdf $f_X(x)$.
 $x \to a^+$ denotes that $x$ tends to $a$ from above.
  $\smallO{\cdot}$ denotes the little-o
notation~\cite{Har58}. $\approx$ denotes approximate equality.

\section{System Model and Problem Statement} \label{sec: sys}
\subsection{System Setup}
Consider a SIMO system in which the transmit  symbol $s$ is drawn uniformly from an $M$-PSK constellation defined as
\begin{align}
\setS_M \triangleq \{e^{j\br{\frac\pi4 + \frac{2\pi }Mi}}: i=0, \ldots, M-1\}.\nonumber
\end{align}
Assuming a receiver equipped with ${N_\textup{r}}$ antennas, the  \black{unobservable analog} received signal prior to quantization is given by
\begin{align}
\y \triangleq \sqrt{\rho}\h s +\n, \label{eqn: sys_model_original}
\end{align}
where  $\h = [h_1, \ldots, h_{N_\textup{r}}]\Trans$ and $\n = [n_1,  \ldots, n_{N_\textup{r}}]\Trans$ represent the channel and additive white Gaussian noise (AWGN) vectors over the $ N_\textup{r} $ receive antennas, respectively. The channel and AWGN vectors are independent, and both are assumed to follow the same distribution $\mathcal{CN}(\mathbf{0}, \mathbf{I}_{N_\textup{r}})$. In this setting, the transmit power $\rho$ represents the transmit SNR.

The received signal $\y$ in~\eqref{eqn: sys_model_original} is further  $n$-bit phase-quantized. Specifically, let $\mathscr{Q}_{n}:\complex \to \setS_{2^n}$ denote the memoryless $n$-bit phase quantization function~\cite{Wu_Liu_et_al_2023}

\begin{align}
\qnbitp{n}{x} \triangleq s^\star\in \argmin_{s\in\setS_{2^n}}\ |s-x|.\label{eqn: Qn_def}
\end{align}
 Due to the above quantization, each $y_i$ is mapped to one of the $2^n$-PSK constellation points. Accordingly, by applying $\mathscr{Q}_n$ element-wise, the \black{observable} quantized received vector is given by
\begin{align}
\r & \triangleq \qnbitp{n}{\y}\in\setS_{2^n}^{N_\textup{r}}. \label{eqn: Q-nb_first_use}
\end{align}
In particular, for $n=2$, we have
\begin{align}
\qnbitp{2}{x} \triangleq \frac1{\sqrt2}\big(\sgn\br{\re(x)} + j\sgn\br{\im(x)}\big),\nonumber
\end{align}
which precisely corresponds to the widely known $1$-bit ADC model in the literature for quantizing a complex value, with one bit for the sign of the real part and another for the sign of the imaginary part~\cite{Y_Li_et_al_BLMMSE, Atzeni_2022}.
In the other extreme, the infinite-bit phase quantizer simplifies to 
\begin{align}
    \qnbitp{\infty}{x} \triangleq \frac x{|x|},\ x\ne0.\nonumber
\end{align}


\subsection{Signal Reception} \label{sec: det}

We first consider coherent detection with perfect CSIR.
\black{We use MRC 
as in~\cite{Jac17, Y_Li_et_al_BLMMSE, Atzeni_2022}
 followed by minimum-distance detection~\cite{Ber22} to obtain the detected symbol as
\begin{align}
\widehat{s}_\mrc\triangleq\qnbitp{m}{\h\Herm\r},\label{eqn: MRC_original}
\end{align}
where $m=\log_2M$ corresponds to the $M$-PSK modulation at the transmitter.} In addition to MRC, SC has also been investigated, e.g.,
in~\cite{Gay21}, where the selection criterion differs from its counterpart
in unquantized systems. To complement existing results, a quantized SIMO-SC system is also considered
in the subsequent analysis.

Due to the quantization at the receiver, it is essential to study
the effect of phase-quantized CSIR. Under such limited
CSIR conditions, signal detection is performed using a majority-decision rule~\cite{Van01}.

\subsection{Diversity and Coding Gains}\label{sec: div_cod}

The diversity gain $G_{\dg}$ and coding gain $G_{\cg}$ are important
metrics that characterize system reliability at high
SNR~\black{\cite[Eq.~(1)]{Gia03}, \cite{Proakis_Dig_comm}} as
\begin{align}
\SEPG \approx (G_{\cg} \rho)^{-G_{\dg}}, \ \rho\to\infty,\nonumber
\end{align}
where 
\begin{align}\SEPG \triangleq \Exp{\Prob{\hat s\ne s|\h}}\label{eqn: sep}\end{align} denotes the average SEP, $s$ is the transmit symbol, $\hat s$ the detected symbol, and $\h$ the instantaneous fading realization. The inner probability is taken with respect to $\n$ and $s$, while the outer expectation is taken with respect to $\h$. The relevance of these metrics has been demonstrated, e.g., in~\cite{Gia03, Ribeiro_Cai_Giannakis_2005, Gay17, Gay21,
Wu_Liu_et_al_2023}. Unlike~\cite{Gayan_2020_Open_J, Gay21, Wu_Liu_et_al_2023},
our approach builds on the insights of~\cite[Prop.~1]{Gia03} and focuses on the important 
channel statistics that determine the diversity and coding gains \black{to quantify the impact of $n$, $M$, and $N_\textup{r}$ on the average SEP performance}.
For completeness,~\cite[Prop.~1]{Gia03} is included in Appendix~\ref{app: dc}.

\subsection{Problem Statement}\label{sec: prob_statement}

The main obstacle in analyzing the SEP of quantized SIMO-MRC systems arises
from the loss of separability between signal and noise after quantization.  For
example, to characterize the high-SNR behavior, the conditional SEP typically
takes the form of $\qfunc{\sqrt{\rho V}}$, where $V$ is a channel-dependent
random variable (see ~Appendix~\ref{app: dc}). However, as mentioned earlier,
conventional analytical methods do not apply to quantized systems in general.
In the following, we introduce a new approach that facilitates tractable SEP
analysis of such systems, thereby extending prior results. \black{In addition, we
examine how the system parameters $n$, $M$, and $N_\textup{r}$ impact the
random variable $V$ to influence the reliability characterization of a
quantized system.}

\section{A Novel Approach to SEP Analysis of Phase-Quantized SIMO-MRC} \label{sec: main}

A conventional SEP analysis with coherent MRC detection requires deriving
the conditional SEP based on~\eqref{eqn: MRC_original} for a given $\h$, followed by averaging over the channel statistics. However, since $\r$ in~\eqref{eqn: MRC_original} is quantized, obtaining the conditional SEP in closed form is
not straightforward. To overcome this difficulty, we propose a new
method that jointly accounts for the randomness of the channel and AWGN by exploiting their inherent circular symmetry.

Recalling~\eqref{eqn: sys_model_original},~\eqref{eqn: Q-nb_first_use}, and~\eqref{eqn: MRC_original}, for a given
$s\in\setS_M$, define the error event set
\begin{align}
\setE_1&\triangleq\mleft\{(\h, \n): \qnbitp{m}{\h^\herm\r}\neq s\mright\}\nonumber
\\&=\mleft\{(\h, \n): \qnbitp{m}{\h^\herm\qnbitp{n}{\y}}\neq s\mright\} \nonumber
    \\&=\mleft\{(\h, \n): \qnbitp{m}{\h^\herm\qnbitp{n}{\sqrt{\rho}\h s +\n}}\neq s\mright\} \label{eqn: original_MRC_events}.
    \end{align} 
Define the indicator function
\begin{align}
\mathbb{I}_{\mathcal{Z}}(z)\triangleq\begin{cases}
        1, & z\in\mathcal{Z},
        \\0, &\text{otherwise}.
    \end{cases} \nonumber
\end{align}
Recalling~\eqref{eqn: sep}, the average SEP over fading and AWGN obtained by the MRC detection in~\eqref{eqn: MRC_original} can be expressed as
\begin{align}
\SEPG &\triangleq \sum_{s\in\setS_M}\Exp{\mathbb{I}_{\setE_1}(\h, \n)|s}\Prob{s}\nonumber
                  \\&= \Exp{\mathbb{I}_{\setE_1}(\h, \n)|s}, \label{eqn: original_MRC_SEP}
\end{align}
where~\eqref{eqn: original_MRC_SEP} holds for any $s\in \setS_M$ due to the symmetry of $\setS_M$ and the assumption of equiprobable input symbols.

Crucial for the forthcoming analysis is the definition of the error event set
\begin{align}
\setE_2&
\triangleq\mleft\{(\h, \n): \qnbitp{m}{
\br{\qnbitp{n}{\h }}^{\herm}\y}\neq s \mright\},\label{eqn: AO_MRC_events}
\end{align}which can be interpreted as the collection of fading and AWGN realizations that lead to error events  \black{in an \textit{auxiliary} SIMO-MRC system employing $n$-bit phase-quantized CSI at each receive antenna, while the received signal vector $\y$ remains unquantized. This auxiliary system is analytically more tractable than the original quantized system.
As shown next, the two systems share the same average SEP, enabling the analysis of the original system through the auxiliary system.}

\begin{theorem}\label{theorem: equivalence_MRC_AO}
Under the assumptions given in the system model, the following holds:
\begin{align}
    \Exp{\mathbb{I}_{\mathcal {E}_1}(\h, \n)|s}=\Exp{\mathbb{I}_{\mathcal {E}_2}(\h, \n)|s},\ \forall s\in\setS_M \nonumber.
\end{align}
\end{theorem}
\begin{IEEEproof}
We have
\begin{align}
\Exp{\mathbb{I}_{\setE_1}(\h, \n)|s}
=\ &\Prob{\qnbitp{m}{\h^\herm \qnbitp{n}{\y}} \ne s|s}\nonumber\\
=\ &\Prob{\qnbitp{m}{\sqrt{\rho + 1}\h^\herm \qnbitp{n}{\y}} \ne s|s}\label{eqn: psi},
\end{align}
which follows from
\begin{align}
\qnbitp{n}{kx} = \qnbitp{n}{x},\ k>0.\label{eqn: si}
\end{align}
From~\eqref{eqn: sys_model_original} and the related system assumptions, it is clear that
\black{$\y$ conditioned on $s$ is distributed as $\setC\setN(\0, \br{\rho + 1}\I_{N_\textup{r}})$ due to the relations
\begin{align}
\Exp{\y|s} & = \sqrt{\rho}\Exp{\h}s + \Exp{\n} = \0,\nonumber \\
\Exp{\y\y^\herm|s} & = \rho\Exp{\h\h^\herm}|s|^2 + \Exp{\n\n^\herm} = (\rho+1)\I_{N_\textup{r}}.\nonumber
\end{align}}
Therefore, both random vectors $[\sqrt{\rho + 1}\h^\tran \quad
\y^\tran]^\tran$ and
$[\y^\herm \quad \sqrt{\rho + 1}\h^\herm]^\tran$ \black{conditioned on $s$} are zero-mean circularly symmetric complex Gaussian with identical covariance matrix given by
\[
\begin{bmatrix}\br{\rho + 1}\I_{N_\textup{r}} & s^*\sqrt{{\rho}\br{\rho + 1}}\I_{N_\textup{r}} \\ s\sqrt{{\rho}\br{\rho + 1}}\I_{N_\textup{r}} & \br{\rho+1}\I_{N_\textup{r}} \end{bmatrix}.
\]
The probability in~\eqref{eqn: psi} is taken with respect to the joint random vector $[{\sqrt{\rho + 1}}\h^\tran \quad {\y^\tran}]^\tran$, which can be replaced by the identically distributed joint random vector $[{\y^\herm} \quad {\sqrt{\rho + 1}}\h^\herm]^\tran$ to yield
\begin{align}
\Exp{\mathbb{I}_{\setE_1}(\h, \n)|s}
=\ &\Prob{\qnbitp{m}{\br{\y^*}^\herm \qnbitp{n}{\sqrt{\rho + 1}\h^*}} \ne s|s}\nonumber.
\end{align}
Invoking~\eqref{eqn: si} and noting that $\qnbitp{n}{\h^*}=(\qnbitp{n}{\h})^*$ holds everywhere except at the boundaries of the phase-quantization function, we obtain
\begin{align}
\Exp{\mathbb{I}_{\setE_1}(\h, \n)|s}
=\ &\Prob{\qnbitp{m}{(\y^*)^\herm (\qnbitp{n}{\h})^*} \ne s|s}\nonumber\\
=\ &\Prob{\qnbitp{m}{\br{\qnbitp{n}{\h}}^\herm \y} \ne s|s}\nonumber\\
=\ &\Exp{\mathbb{I}_{\setE_2}(\h, \n)|s}, \ \forall s\in\setS_M.\nonumber
\end{align}
\end{IEEEproof}

Theorem~\ref{theorem: equivalence_MRC_AO} allows us to determine the average SEP for
a phase-quantized SIMO-MRC system over i.i.d. Rayleigh fading in~\eqref{eqn: original_MRC_SEP} using the alternative characterization in \black
{the auxiliary system} in~\eqref{eqn: AO_MRC_events}, i.e., 
\begin{align}
    \SEPG = \Exp{\mathbb{I}_{\mathcal {E}_2}(\h, \n)|s}. \label{eqn: MRC_SEP_alternative}
\end{align} 
The right-hand side (RHS) of~\eqref{eqn: MRC_SEP_alternative} which pertains to the auxiliary system
lends itself to convenient analysis, unlike the RHS of~\eqref{eqn:
original_MRC_SEP}. 

More specifically, let us write
$
h_i=|h_i|e^{j\theta_i}
$. Due to i.i.d. Rayleigh fading, we have
\begin{align}|h_i|^{2} \sim \exp(1), \quad\theta_i \sim \setU(-\pi, \pi), \label{eqn: Rayleigh_fading_basic}
\end{align} where  $|h_i|$ and $\theta_i$ are independent, for $i=1,
\ldots, {N_\textup{r}}$.
\black{Now, recall~\eqref{eqn: Qn_def} and define
\begin{align}
\widetildetheta_i\triangleq\arg\br{\qnbitp{n}{h_i^*}h_i} \sim \setU\mleft(-\frac{\pi}{2^n}, \frac{\pi}{2^n}\mright),\label{eqn: E2_interp_2}
\end{align}
where $\widetildetheta_i$ is independent of $|h_i|$, for $i=1, \ldots, {N_\textup{r}}$.
Furthermore, let $\phi_i \triangleq \theta_i - \widetildetheta_i$ be the quantized phase of the antenna branch $i$, which allows us to write
$\qnbitp{n}{h_i} = e^{j\phi_i}$.}
Further, define
\begin{alignat}{2}
y_\summa &\triangleq \frac{1}{\sqrt{N_\textup{r}}}\br{\qnbitp{n}{\h}}\Herm\y =\sqrt{\frac\rho{N_\textup{r}}}(\qnbitp{n}{\h})\Herm\h s + n_\summa\nonumber
\\       &=  \frac{1}{\sqrt{N_\textup{r}}}\sum_{i=1}^{N_\textup{r}} \widetilde{y}_i
= \sqrt{\frac\rho{N_\textup{r}}}\sum_{i=1}^{N_\textup{r}} \widetilde{h}_is + n_\summa, \label{eqn: E2_interp_7}
\end{alignat}
with
$\widetilde{y}_i \triangleq e^{-j\phi_i} y_i$,
$\widetilde{n}_i \triangleq e^{-j\phi_i}n_i$, and
$\widetilde{h}_i\triangleq|h_i| e^{j\widetilde\theta_i}$.
Since the phase rotation does not change the AWGN distribution, we have $\widetilde{n}_i, n_i \sim \mathcal{CN}(0, 1)$, for $i=1, \ldots, N_\textup{r}$, and therefore we have
$
n_\summa \triangleq \sum_{i=1}^{N_\textup{r}}\frac{\widetilde{n}_i}{\sqrt N_\textup{r}} \sim \mathcal{CN}(0, 1)
$.
Finally, defining
\begin{align}
\widetilde{\setE}_2&\triangleq
\mleft\{(\widetilde{h}_i \ \forall i, n_\summa): \qnbitp{m}{y_\summa} \neq s\mright\} \nonumber
\end{align}
allows us to obtain the relation
\begin{align}
\Exp{\mathbb{I}_{\setE_2}(\h, \n)|s}=\Exp{\mathbb{I}_{\widetilde{\setE}_2}(\widetilde{h}_i\,  \forall i, n_\summa)|s}.\label{eqn: E2_tildeE2}
\end{align}
Through the above steps, the average SEP analysis of a SIMO-MRC
system under phase quantization has been transformed into an equivalent single-channel problem (\emph{cf}.~\eqref{eqn: E2_interp_7}), with the equivalent channel gain given by $
\frac{1}{\sqrt N_\textup{r}}\sum_{i=1}^{N_\textup{r}}\widetilde{h}_i
$.

\black{
\section{MISO-SIMO Duality and Average SEP Analysis with \texorpdfstring{$M$}{M}-PSK Modulation} \label{sec: duality}
}
Theorem~\ref{theorem: equivalence_MRC_AO} enables us to analyze the
average SEP for a general SIMO case based on the auxiliary system pertaining to $\setE_2$.
From~\eqref{eqn:
E2_interp_7}--\eqref{eqn: E2_tildeE2}, we have 
\begin{align}
\SEPG=\Exp{\mathbb{I}_{\widetilde{\setE}_{2}}(\widetilde h_i\, \forall i, n_\summa)|s}\label{eqn: general_equiv_3}.
\end{align}
It turns out that the average SEP of the auxiliary system is further equivalent to the average SEP of  the MISO-MRT system in~\cite{Wu_Liu_et_al_2023}, as will be shown below.

In~\cite{Wu_Liu_et_al_2023}, the SEP performance for a MISO system with
low-resolution DACs and perfect CSI at the transmitter (CSIT) was investigated. In this setting, the received signal is modeled as 
\begin{align}
y \triangleq \sqrt{\frac\rho {N_\textup{t}}}\h^\tran\qnbitp{n}{\h^*s} + \widetilde{w} \in \complex,\label{eqn: MISO system model}
\end{align}
where $\h \sim \mathcal{CN}(\0, \I_{N_\textup{t}})$ denotes the MISO channel,
$\qnbitp{n}{\h^*s}$ is the normalized MRT quantized
constant envelope (QCE) transmitted symbol vector, the symbols $s \in \setS_M$
are drawn with equal probability, $\widetilde{w} \sim \mathcal{CN}(0, 1)$ is the AWGN, $\rho$
represents the transmit SNR, and $N_\textup{t}$ is the number of transmit antennas. At the
receiver, the detected symbol is obtained as $\hat{s} \triangleq \qnbitp{m}{y}$.
The SEP corresponding the above MISO system is
\begin{equation}
\begin{aligned}
\SEPG^\miso = \Exp{\mathbb{I}_{\setE^\miso}(\h, \widetilde w)|s},
\end{aligned}
\end{equation}
with 
\begin{equation}
\begin{aligned}
\setE^\miso \triangleq \{(\h, \widetilde{w}): \qnbitp{m}{y} \ne s\}.
\end{aligned}
\end{equation}

\begin{proposition} \label{theo: simo_miso_duality_ASER}
Assume $N_\textup{t}=N_\textup{r}$. When the  number of bits used for quantization and  the modulation order are identical, a MISO system with low-resolution
DACs and perfect CSIT is dual to a SIMO system with
low-resolution ADCs and perfect CSIR,  in the sense that both systems exhibit identical average SEP performance. 
\end{proposition}

\begin{IEEEproof}
The MISO system described in~\eqref{eqn: MISO system model} is equivalent to 
\begin{align}
y &= \sqrt{\frac\rho {N_\textup{t}}}(\h^* s)\Herm\qnbitp{n}{\h^*s} s + \widetilde{w}  \nonumber
\\&=\sqrt{\frac\rho {N_\textup{t}}}\h_\miso\Herm\qnbitp{n}{\h_\miso} s + \widetilde{w} \label{eqn: MISO_sys_model_equiv}
\end{align}
where $\h_\miso\triangleq\h^*s$ satisfies $\h_\miso \sim
\mathcal{CN}(0, \I_{N_\textup{t}})$ and is independent of the AWGN $\widetilde{w}$.
Comparing~\eqref{eqn: E2_interp_7} and~\eqref{eqn: MISO_sys_model_equiv}
leads to the conclusion that, with $N_\textup{t} = N_\textup{r}$, these two systems are equivalent in the average SEP sense. As a result, given the same assumption on the number of quantization bits and the modulation
order, the subsequent detection performance based on~\eqref{eqn:
E2_interp_7} and~\eqref{eqn: MISO_sys_model_equiv} are the same.
\end{IEEEproof}

\noindent  Based on the duality in Proposition~\ref{theo: simo_miso_duality_ASER}, all the results derived for a SIMO-MRC system in Section~\ref{sec: qpsk} are valid for the corresponding dual MISO
system considered in~\cite{Wu_Liu_et_al_2023}.


With $M$-PSK modulation, an exact average SEP expression in the form of  expectation of a Gaussian $Q$-function does not seem to be available~\cite{Gia03, Wu_Liu_et_al_2023}. Instead, upper- and lower-bounds on the SEP can be obtained~\cite[p. 320, Problem 5.5]{Stuber_mobile_comm}\cite{Wu_Liu_et_al_2023}, which suffice for deriving the diversity gain. Specifically, based on \eqref{eqn: E2_interp_7}, the SEP of an $M$-PSK-modulated, $n$-bit phase-quantized SIMO-MRC system can be bounded as
\begin{align}
    \SEPG_\lb\leq\SEPG\leq 2\SEPG_\lb, \label{eqn: MPSK_SEP_bounds}
\end{align}
with $\SEPG_\lb \triangleq \Exp{\qfunc{\sqrt{\rho}\widetilde T}}$,
\begin{align}
    \widetilde T \triangleq \frac{1}{\sqrt{N_\textup{r}}}\sum_{i=1}^{N_\textup{r}}\widetilde Z_i, \label{eqn: MPSK_key_stat}
\end{align}
$ \widetilde Z_i \triangleq \sqrt{2}|h_i|\sin\br{\widetilde\theta_i + \frac{\pi}{M}}$, $|h_i|^2 \sim \exp(1)$, and $\widetildetheta_i \sim \setU\br{-\frac{\pi}{2^n}, \frac{\pi}{2^n}}$.
Note that by replacing $N_r$ with the number of transmit antennas in~\cite{Wu_Liu_et_al_2023}, the statistic in \eqref{eqn: MPSK_key_stat} is similar but still not identical to \cite[(13)]{Wu_Liu_et_al_2023}. In light of the duality in Proposition~\ref{theo: simo_miso_duality_ASER}, similar analyses to those in~\cite{Wu_Liu_et_al_2023} can be applied to \eqref{eqn: MPSK_SEP_bounds}--\eqref{eqn: MPSK_key_stat} here to obtain the following corollary. 

\begin{corollary}\label{cor: general_diversity}
With perfect CSIR, an $M$-PSK-modulated, $n$-bit phase-quantized SIMO-MRC system
achieves the  diversity gain
\begin{align}\label{eqn: general_diversity_gain}
    G_{\dg, \mrc}=\begin{cases}
        0, &2^n<M,\\
        \frac{N_\textup{r}}{2},  & 2^n = M,\\
        N_\textup{r}, & 2^n > M.
    \end{cases}
\end{align}
\end{corollary}
{\color{black}Clearly, Corollary~\ref{cor: general_diversity} is dual to~\cite[Thm.~2]{Wu_Liu_et_al_2023}.} With $M$-ary modulation at the transmitter, Corollary~\ref{cor: general_diversity}
reveals \black{two distinct phase transitions in the diversity gain
around $2^n = M$ which has been previously observed in~\cite{Gayan_2020_Open_J, Gay21,
Wu_Liu_et_al_2023}. Therefore, the design guidelines proposed in \cite{Wu_Liu_et_al_2023} for a MISO system can be adopted for a $M$-PSK modulated, $n$-bit phase-quantized SIMO system, namely, the system should satisfy $M<2^n$  in order to achieve the full diversity gain or the system should satisfy $M=2^n$ to at least achieve a vanishing SEP. More on phase transitions and design criteria will be discussed for a QPSK-modulated system in Section~\ref{sec: qpsk}}. 

\black{
\begin{proposition}\label{prop: coding_general}
    Based on \eqref{eqn: MPSK_SEP_bounds}, the coding gains of $M$-PSK-modulated, $n$-bit phase-quantized SIMO-MRC system for $M=2^n$ and $M < 2^n$ are
\begin{subequations}\label{eqn: general_coding_gain}
    \begin{align}
        G_{\cg, \mrc}^\equalcase &= \br{\frac{M^{N_\textup{r}}2^{-{N_\textup{r}}-1}k_{n, M, N_\textup{r}}{N_\textup{r}}^{\frac{N_\textup{r}}2}}{\pi^{\frac{{N_\textup{r}}+1}{2}}{N_\textup{r}}!}\gammaf{\frac{{N_\textup{r}}+1}{2}}}^{-\frac2{N_\textup{r}}},\nonumber\\
        G_{\cg, \mrc}^\inequalcase &= \br{\frac{2^{n{N_\textup{r}}-1}k_{n, M, N_\textup{r}}N_\textup{r}^{N_\textup{r}}}{{\pi}^{N_\textup{r} + \frac12}(2{N_\textup{r}})!}\gammaf{{N_\textup{r}} + \frac12}\mright.\nonumber\\&\phantom=\mleft.\times\br{\cot\br{\frac\pi M - \frac\pi{2^n}} - \cot\br{\frac\pi M + \frac\pi{2^n}}}^{N_\textup{r}}}^{-\frac1{N_\textup{r}}},\nonumber
    \end{align}
\end{subequations}
    respectively, up to a scaling factor $k_{n, M , N_\textup{r}}$ bounded between $1$ and $2$.
\end{proposition}
\begin{IEEEproof}
See Appendix~\ref{app: coding_general}.
\end{IEEEproof}
}

The duality in Proposition~\ref{theo: simo_miso_duality_ASER} and the bounds in \eqref{eqn: MPSK_SEP_bounds}--\eqref{eqn: general_coding_gain} are exemplified numerically in
Fig.~\ref{fig: miso-simo}, which shows identical SEP curves for the MISO and SIMO systems for $N_\textup{r} \in \{2, 8\}$, $M=8$, and $n\in\{3, 4\}$. 
\begin{figure}[!t]
\centering
\begin{tikzpicture} [auto]
\begin{semilogyaxis}
[
width=8cm,
height=6cm,
xmin=0, xmax=20,
ymin=10^-8, ymax=10^0,
ytick={10^-8, 10^-6,10^-4,10^-2,10^0},
xlabel = {$\rho$ [dB]},
ylabel = {Average SEP},
ylabel near ticks,
x label style={font=\scriptsize},
y label style={font=\scriptsize},
ticklabel style={font=\scriptsize},
legend style = {font=\scriptsize},
legend pos = south west,
grid=both,
major grid style={line width=.2pt,draw=gray!30},
mark options = {solid},
]

\addplot    [thin, dotted, mark=o, mark repeat=4] table [ y=ser, x=rho, col sep=comma ] {data/misoNr2M8n3.txt};\addlegendentry{MISO (Sim.)}
\addplot    [thin, dashed, mark=x, mark repeat=4] table [ y=mrc, x=rho, col sep=comma ] {data/simoNr2M8n3Wu.txt};\addlegendentry{SIMO (Sim.)}
\addplot    [thin, dashed, mark=triangle, mark repeat=4] table [ y expr=2*\thisrow{wu}, x=rho, col sep=comma ] {data/simoNr2M8n3Wu.txt};\addlegendentry{Upper bound \eqref{eqn: MPSK_SEP_bounds}}
\addplot    [thin, solid] table [ y=bound, x=rho, col sep=comma ] {data/simoNr2M8n3Wu.txt};\addlegendentry{Bound \eqref{eqn: general_diversity_gain}--\eqref{eqn: general_coding_gain}}

\addplot    [thin, dotted, mark=o, mark repeat=4] table [ y=ser, x=rho, col sep=comma ] {data/misoNr8M8n3.txt};
\addplot    [thin, dashed, mark=x, mark repeat=1] table [ y=mrc, x=rho, col sep=comma ] {data/simoNr8M8n3Wu.txt};
\addplot    [thin, dashed, mark=triangle, mark repeat=1] table [ y expr=2*\thisrow{wu}, x=rho, col sep=comma ] {data/simoNr8M8n3Wu.txt};
\addplot    [thin, solid] table [ y=bound, x=rho, col sep=comma ] {data/simoNr8M8n3Wu.txt};

\addplot    [thin, dotted, mark=o, mark repeat=4] table [ y=ser, x=rho, col sep=comma ] {data/misoNr2M8n4.txt};
\addplot    [thin, dashed, mark=x, mark repeat=1] table [ y=mrc, x=rho, col sep=comma ] {data/simoNr2M8n4Wu.txt};
\addplot    [thin, dashed, mark=triangle, mark repeat=1] table [ y expr=2*\thisrow{wu}, x=rho, col sep=comma ] {data/simoNr2M8n4Wu.txt};
\addplot    [thin, solid] table [ y=bound, x=rho, col sep=comma ] {data/simoNr2M8n4Wu.txt};

\addplot    [thin, dotted, mark=o, mark repeat=4] table [ y=ser, x=rho, col sep=comma ] {data/misoNr8M8n4.txt};
\addplot    [thin, dashed, mark=x, mark repeat=1] table [ y=mrc, x=rho, col sep=comma ] {data/simoNr8M8n4Wu.txt};
\addplot    [thin, dashed, mark=triangle, mark repeat=1] table [ y expr=2*\thisrow{wu}, x=rho, col sep=comma ] {data/simoNr8M8n4Wu.txt};
\addplot    [thin, solid] table [ y=bound, x=rho, col sep=comma ] {data/simoNr8M8n4Wu.txt};

\node[text opacity=1, anchor=north west, fill=white, fill opacity=.4] at (13, 4*10^-1) {\scriptsize $N_\textup{r}=2,\ n=3$};
\node[text opacity=1, anchor=north west, fill=white, fill opacity=.4] at (13, 9*10^-3) {\scriptsize $N_\textup{r}=2,\ n=4$};
\node[text opacity=1, anchor=north west, fill=white, fill opacity=.4] at (13, 4*10^-4) {\scriptsize $N_\textup{r}=8,\ n=3$};
\node[text opacity=1, anchor=north west, fill=white, fill opacity=.4] at (13, 9*10^-7) {\scriptsize $N_\textup{r}=8,\ n=4$};

\end{semilogyaxis}
\end{tikzpicture}
\caption{MISO-SIMO duality, with $N_\textup{r}=N_\textup{t} \in \{2, 8\}$, $8$-PSK,
and \black{$n\in\{3, 4\}$.}}
\label{fig: miso-simo}
\end{figure}

\black{
Corollary~\ref{cor: general_diversity} implies an error floor with $M > 2^n$ which can be lower bounded as follows:
\begin{subequations}
\begin{align}
\SEPG^\efcase &\ge \SEPG_\lb = \Exp{\qfunc{\sqrt{\rho}\widetilde T}}\ge \int_{-\infty}^0\qfunc{\sqrt{\rho}\widetilde T}f_{\widetilde T}(t)\, dt \nonumber
\\            &\ge \frac12 \Prob{\widetilde T < 0} = \frac12 \Prob{\sum_i \widetilde Z_i < 0}\label{eqn: ef1}
\\            &\ge  \frac12\Prob{\widetilde Z_i < 0,\ \forall i} = \frac12\br{\Prob{Z_1 < 0}}^{N_\textup{r}} \label{eqn: ef2}
\\            &=  \frac12 \br{\frac12\br{1-\frac{2^n}{M}}}^{N_\textup{r}},\label{eqn: ef}
\end{align}
\end{subequations}
where $f_{\widetilde T}(t)$ is the pdf of $\widetilde T$. The inequality in~\eqref{eqn: ef1} follows from
$\qfunc{x} \ge \qfunc{0} = \frac12$, for $x\le0$, whereas the equality in \eqref{eqn: ef2} follows from $\widetilde Z_i$'s being i.i.d. We have used $\widetilde\theta_1\sim\setU\br{-\frac{\pi}{2^n}, \frac{\pi}{2^n}}$ in \eqref{eqn: ef}.  
The presence of an error floor for $M > 2^n$ stems from the fact that, even in the absence of
AWGN, each quantized observation may correspond to more than one transmit symbol since $\frac{M}{2^n}>1$, as also discussed in \cite[Thm.~1]{Gayan_2020_Open_J}.
}

\section{Average SEP Analysis with QPSK Modulation} \label{sec: qpsk}

In this section, we analyze the average SEP for a QPSK-modulated, low-resolution
phase-quantized SIMO system. Both SIMO-MRC and SIMO-SC architectures are considered, with the main focus on the MRC case. \black{While the coding gains characterized in Proposition~\ref{prop: coding_general} are not exact, the exact SEP expression for a QPSK-modulated system allows derivation of exact coding gains. In addition, the QPSK-modulated system provides additional insight into the phase transitions revealed in Corollary~\ref{cor: general_diversity}.}

\subsection{Exact Results on a SIMO-MRC System}\label{sec: mrc}

We first obtain an exact SEP expression based on~\eqref{eqn:
MRC_SEP_alternative}--\eqref{eqn: E2_interp_7}. Following the discussion of~\eqref{eqn: original_MRC_SEP}, without loss of generality, we
analyze the average SEP for QPSK modulation (i.e., $M=4$, $m=2$) by fixing the transmitted symbol to $s=e^{j\frac\pi4} \in \setS_M$. In this case, we have
\begin{equation}\nonumber
\begin{aligned}
\re(y_\summa) &= \sqrt{\frac{\rho}{N_\textup{r}}} \sum_{i=1}^{N_\textup{r}} \abs{h_i}\cos\mleft(\widetildetheta_i + \frac\pi4\mright) + \re\mleft(n_\summa\mright),\\
\im(y_\summa) &= \sqrt{\frac{\rho}{N_\textup{r}}} \sum_{i=1}^{N_\textup{r}} \abs{h_i}\sin\mleft(\widetildetheta_i + \frac\pi4\mright) + \im\mleft(n_\summa\mright).
\end{aligned}
\end{equation}
Given perfect CSIR, the average SEP is given by
\begin{align}
\SEP        =\ & 1 - \Prob{\re(y_\summa) > 0, \im(y_\summa) > 0}\nonumber
          \\=\ & 1- \Exp{\qfunc{-\sqrt{\frac{\rho}{N_\textup{r}}}\sum_{i=1}^{N_\textup{r}} \sqrt2|h_i|\cos\mleft(\widetildetheta_i+\frac\pi4\mright)}\mright. \nonumber
\\  \phantom=\ & \mleft.\times\qfunc{-\sqrt{\frac{\rho}{N_\textup{r}}}\sum_{i=1}^{N_\textup{r}} \sqrt2|h_i|\sin\mleft(\widetildetheta_i+\frac\pi4\mright)}} \nonumber
          \\=\ & 1- \Exp{\qfunc{-\sqrt{\rho}T}\qfunc{-\sqrt{\rho}\widetilde T}} \nonumber
\\=\ & 2\Exp{\qfunc{\sqrt{\rho}T}} - \Exp{\qfunc{\sqrt{\rho}T} \qfunc{\sqrt{\rho}\widetilde T}},\label{eqn: MRC_theorem_proof_2}
\end{align}
where we have defined the following key statistics:
\begin{subequations}\label{eqn: keystat1}
\begin{align}
   Z_i             &\triangleq \sqrt2|h_i|\cos(\widetildetheta_i+\frac\pi4) = |h_i|(\cos\widetildetheta_i - \sin\widetildetheta_i),\label{eqn: Z_def}
\\ \widetilde{Z}_i &\triangleq \sqrt2|h_i|\sin(\widetildetheta_i+\frac\pi4) =  |h_i|(\cos\widetildetheta_i + \sin\widetildetheta_i),\nonumber
\\T                &\triangleq\frac1{\sqrt {N_\textup{r}}}\sum_{i=1}^{N_\textup{r}} Z_i, \quad \widetilde{T} \triangleq  \frac1{\sqrt {N_\textup{r}}}\sum_{i=1}^{N_\textup{r}} \widetilde{Z}_i.\label{eqn: T_def}
\end{align}
\end{subequations}
\black{Observe that \eqref{eqn: T_def} is consistent with \eqref{eqn: MPSK_key_stat}}.
Based on Remark~\ref{rem: ident} in Appendix~\ref{app: key_stats}, we use the fact that $T$ and $\widetilde T$ in~\eqref{eqn: MRC_theorem_proof_2} are identically
distributed.
Further define
\begin{align}
U \triangleq T^2,\quad \widetilde U \triangleq \widetilde T^2,\label{eqn: U_def}
\end{align}
which will be used later. The properties, relations, and selected
pdfs of the random variables in~\eqref{eqn: keystat1}--\eqref{eqn: U_def} are summarized in Appendix~\ref{app: key_stats}.

Since we have $\widetildetheta_i\sim \setU\mleft( -\frac\pi{2^n}, \frac\pi{2^n}\mright)$, the distributions of
the random variables in~\eqref{eqn: keystat1}--\eqref{eqn: U_def} depend explicitly on the quantization resolution $n$. 
Consequently, the average SEP also varies with $n$. In the following, we analyze the average SEP behavior for different values of $n$.

\subsubsection{The case with \texorpdfstring{$n=1$}{n=1}}
\black{
Based on \eqref{eqn: ef}, with $n=1,\ M=4$, we have
\begin{align}
    \SEPG^\onebit \ge 2^{-1 -2N_\textup{r}}\nonumber
\end{align}
as the lower bound to the non-vanishing error floor.
}

\subsubsection{The case with \texorpdfstring{$n=2$}{n=2}}


\begin{proposition}\label{prop: mrc_ser}
For a QPSK-modulated, $2$-bit phase-quantized SIMO-MRC system with $N_\textup{r}$ receive
antennas under i.i.d. Rayleigh fading, the exact average SEP is
\begin{align}
\SEP^\twobit &= 2\Exp{\qfunc{\sqrt{\rho U}}} - \mleft(\Exp{\qfunc{\sqrt{\rho U}}}\mright)^2\label{eqn: mrc_ser},
\end{align}
where $U$ is defined in~\eqref{eqn: U_def}.
The
corresponding diversity and coding gains are
\begin{subequations}\label{eqn: MRC_high_SNR}
\begin{align}
G_{\dg, \mrc}^\twobit &= \frac{{N_\textup{r}}}{2}, \label{eqn: MRC_diversity_gain}\\
G_{\cg, \mrc}^\twobit &= \mleft(\frac{1}{{N_\textup{r}}!}2^{N_\textup{r}}\pi^{-\frac{N_\textup{r}+1}{2}}N_\textup{r}^{\frac{N_\textup{r}}{2}}\Gamma\mleft(\frac{{N_\textup{r}}+1}2\mright)\mright)^{-\frac2{N_\textup{r}}}, \label{eqn: MRC_high_SNR_coeffi}
\end{align}
\end{subequations}
respectively.
\end{proposition}
\begin{IEEEproof}
\black{
The diversity gain in \eqref{eqn: MRC_diversity_gain} follows directly from Corollary~\ref{cor: general_diversity}. Observing that the SEP at high SNR is dominated by the first term on the RHS of~\eqref{eqn: mrc_ser}, invoke Proposition~\ref{prop: coding_general} with $k_{2, 4, N_\textup{r}}=2$ to obtain the coding gain in \eqref{eqn: MRC_high_SNR_coeffi}.}
\end{IEEEproof}

\begin{figure}[!t]
\centering
\begin{tikzpicture} [auto]
\begin{semilogyaxis}
[
width=8cm,
height=6cm,
ymax=10^0,
ymin=10^(-8),
xmin=0,
xmax=20,
ytick={10^0,10^-2,10^-4,10^-6, 10^-8},
xlabel = {$\rho$ [dB]},
ylabel = {Average SEP},
ylabel near ticks,
x label style={font=\scriptsize},
y label style={font=\scriptsize},
ticklabel style={font=\scriptsize},
legend style = {font=\scriptsize, fill=white, fill opacity=0.6, text opacity=1},
legend pos = south west,
grid=both,
major grid style={line width=.2pt,draw=gray!30},
mark options = {solid},
]

\addplot    [thin, dotted, mark=x, mark repeat=2] table [ y=mrc, x=rho, col sep=comma ] {data/mrcNr1.txt}; \addlegendentry{Simulated}
\addplot    [thin, dashed, mark=o, mark repeat=2] table [ y=exact, x=rho, col sep=comma ] {data/mrcNr1.txt}; \addlegendentry{Theoretical \eqref{eqn: mrc_ser}}
\addplot    [thin, solid]  table [ y=bound, x=rho, col sep=comma ] {data/mrcNr1.txt}; \addlegendentry{Bound \eqref{eqn: MRC_high_SNR}}


\addplot    [thin, dotted, mark=x, mark repeat=2] table [ y=mrc, x=rho, col sep=comma ] {data/mrcNr4.txt};
\addplot    [thin, dashed, mark=o, mark repeat=2] table [ y=exact, x=rho, col sep=comma ] {data/mrcNr4.txt};
\addplot    [thin, solid]  table [ y=bound, x=rho, col sep=comma ] {data/mrcNr4.txt};

\addplot    [thin, dotted, mark=x, mark repeat=2] table [ y=mrc, x=rho, col sep=comma ] {data/mrcNr8.txt};
\addplot    [thin, dashed, mark=o, mark repeat=2] table [ y=exact, x=rho, col sep=comma ] {data/mrcNr8.txt};
\addplot    [thin, solid]  table [ y=bound, x=rho, col sep=comma ] {data/mrcNr8.txt};

\addplot    [thin, dotted, mark=x, mark repeat=2] table [ y=mrc, x=rho, col sep=comma ] {data/mrcNr16.txt};
\addplot    [thin, dashed, mark=o, mark repeat=2] table [ y=exact, x=rho, col sep=comma ] {data/mrcNr16.txt};
\addplot    [thin, solid, domain=0:12] {11.2263959688588 * 10^(-8*x/10)}; 

\node[font=\footnotesize, anchor=north west] at (10.5, 9*10^-1) {$N_\textup{r}=1$};
\node[font=\footnotesize, anchor=north west] at (10.5, 2.5*10^-2) {$N_\textup{r}=4$};
\node[font=\footnotesize, anchor=north west] at (10.5, 3.5*10^-4) {$N_\textup{r}=8$};
\node[font=\footnotesize, anchor=north west] at (10.5, 1*10^-7) {$N_\textup{r}=16$};

\end{semilogyaxis}
\end{tikzpicture}
\caption{Average SEP versus $\rho$ for a QPSK-modulated, $2$-bit phase-quantized SIMO-MRC system, with $N_\textup{r}\in\{1, 4, 8, 16\}$. The corresponding SEP bound is specified by~\eqref{eqn: MRC_high_SNR}.}
\label{fig: mrc}
\end{figure}

\noindent Fig.~\ref{fig: mrc} provides the SEP simulation results of a QPSK-modulated, $2$-bit phase-quantized SIMO-MRC system for $N_\textup{r}\in\{1, 4, 8, 16\}$, which corroborate the above SEP expression as well as its high-SNR characterization. Moreover, based on Remark~\ref{rem: chi2} in Appendix~\ref{app: key_stats},~\eqref{eqn: mrc_ser}
can be simplified to
\begin{align}
\SEP^\twobit &= 1- \br{\Exp{\qfunc{-\sqrt{\frac{\rho}{N_\textup{r}}}\sum_{i=1}^{N_\textup{r}}\sqrt2|\re(h_i)|}}}^2\nonumber
\\           &= 1- \br{\Exp{\qfunc{-\sqrt{\frac{2\rho}{N_\textup{r}}\|\re(\h)\|_1^2}}}}^2.\nonumber
\end{align}

For comparison, the SEP for a QPSK-modulated, unquantized SIMO-MRC system is given by~\cite{Sim00}
\begin{equation}\label{eqn: SEP_unq}
\begin{aligned}
\SEP^\uq = 1 - \Exp{ \br{\qfunc{-\sqrt{\rho\norm{\h}_2^2}}}^2},
\end{aligned}
\end{equation}
with diversity and coding gains given by
\begin{align}
 G_{\dg, \mrc}^{\rm \uq} = N_\textup{r},\quad 
 G_{\cg, \mrc}^{\rm \uq} = 2\ncr{2N_\textup{r}}{N_\textup{r}}^{-\frac{1}{N_\textup{r}}},\nonumber
\end{align}
respectively.
Comparing the above with~\eqref{eqn: MRC_diversity_gain}--\eqref{eqn: MRC_high_SNR_coeffi}, we observe
that the $2$-bit phase-quantized counterpart induces a loss of $\frac{N_\textup{r}}{2}$
in diversity gain. 

\begin{remark}\label{remark: quant_unq_rat}
{\color{black}For the same diversity gain $N_\textup{r}$,  the coding gain of a QPSK-modulated, $2$-bit phase-quantized SIMO-MRC system with $2N_\textup{r}$ receive antennas and that  of its unquantized counterpart with $N_\textup{r}$ receive antennas are related as
\begin{equation}
\begin{aligned}
\frac{G_{\cg, \mrc}^{\rm \twobit}(2N_\textup{r}) }{G_{\cg, \mrc}^{\rm \uq}(N_\textup{r}) } &= \frac \pi{4 {N_\textup{r}}}\mleft[\frac{\br{2N_\textup{r}}!}{N_\textup{r}!}\mright]^{\frac1{N_\textup{r}}} \in \mleft(\frac{\pi}{e}, \frac{\pi}{2}\mright],\label{eqn: quant_unq_rat}
\end{aligned}
\end{equation}
where the lower bound is obtained by applying Stirling's formula to the factorials for large $N_\textup{r}$ and the upper bound is achieved for $N_\textup{r}=1$. Based on \eqref{eqn: quant_unq_rat}, we conclude that, to ensure the SEP achieved by an unquantized SIMO-MRC system at high SNR, it suffices to use twice the number of receive antennas when $2$-bit phase-quantization is applied.}
\end{remark}
For clarity, \eqref{eqn: quant_unq_rat} is also plotted in Fig.~\ref{fig: quant-unq-rat} in logarithmic scale versus $N_\textup{r}$. 
\begin{figure}
\centering
\begin{tikzpicture}[auto]
\begin{axis}[
width=8.0cm,
height=6.3cm,
xmin=0, xmax=50,
ylabel = {Ratio of coding gains [dB]},
xlabel = {$N_\textup{r}$},
ylabel near ticks,
x label style={font=\scriptsize},
y label style={font=\scriptsize},
ticklabel style={font=\scriptsize},
ytick = {.5, 10*log10(pi/e), 1, 1.5, 10*log10(pi/2)},
yticklabels = {$0.5$, $10\log_{10}(\frac\pi e)$, $1$, $1.5$, $10\log_{10}(\frac \pi 2)$}, 
grid=both,
major grid style={line width=.2pt,draw=gray!30},
]
\addplot [
    domain=1:50, 
    samples=50, 
    thin,
]
{10*log10(pi/(4*x) * (factorial(2*x)/factorial(x))^(1/x))};

\end{axis}
\end{tikzpicture}
\caption{Ratio of the coding gains in \eqref{eqn: quant_unq_rat} vs. diversity gain $N_\textup{r}$~\cite{C}.}
\label{fig: quant-unq-rat}
\end{figure}

\subsubsection{The case with \texorpdfstring{$n\ge3$}{n>=3}}

For $n\ge3$, noting that the random variables $T$ and $\widetilde T$ are nonnegative and using Remark~\ref{rem: ident} in Appendix~\ref{app: key_stats}, we obtain the exact
average SEP for a QPSK-modulated, $n$-bit phase-quantized SIMO-MRC system from~\eqref{eqn: MRC_theorem_proof_2}
as
\begin{align}
\SEP^\nbit     &\!=\! 2\Exp{\qfunc{\sqrt{\rho U}}} \!-\! \Exp{\qfunc{\sqrt{\rho U}}\qfunc{\sqrt{\rho \widetilde U}}}
\label{eqn: mrc_ser_n}.
\end{align}

\begin{proposition}\label{prop: mrc_ser_n_high_snr}
For a QPSK-modulated, $n$-bit phase-quantized SIMO-MRC system  with $n \ge 3$ and
$N_\textup{r}$ receive antennas under i.i.d. Rayleigh fading, the diversity and coding gains are given by
\begin{subequations}\label{eqn: MRC_high_SNR_n}
\begin{align}
G_{\dg, \mrc}^\nbit &= N_\textup{r}, \label{eqn: MRC_diversity_gain_n}\\
G_{\cg, \mrc}^\nbit &= \frac{(N_\textup{r}!)^{\frac1{N_\textup{r}}}}{N_\textup{r}}\frac{\pi}{2^{n-1}}\cot\frac\pi{2^{n-1}}, \label{eqn: MRC_high_SNR_coeffi_n}
\end{align}
\end{subequations}
respectively.
\end{proposition}
\begin{IEEEproof}
\black{
This proof follows similar lines to the proof of Proposition~\ref{prop: mrc_ser} and the details are omitted.}
\end{IEEEproof}

In the limit of $n\to\infty$, the phase error in~\eqref{eqn: E2_interp_7} vanishes due to $\widetildetheta_i=0$, for $i=1, \ldots, N_\textup{r}$.  Without loss of generality, consider
$s=e^{j\frac\pi4} \in \setS_4$. In this case, we obtain
\begin{equation}
\begin{aligned}
y_\summa^{\infty} = \sqrt{\frac{\rho}{2N_\textup{r}}} \sum_{i=1}^{N_\textup{r}}\abs{h_i} (1+j)  +n_\summa,\label{eqn: new_y_sum_E2G_n_infinity}
\end{aligned}
\end{equation}
which coincides with a QPSK-modulated, unquantized SIMO system employing equal-gain combining.

\begin{remark}\label{remark: EGC_equiv}
From~\eqref{eqn: new_y_sum_E2G_n_infinity}, the effective decision statistic is $\|\h\|_1 = \sum_{i=1}^{N_\textup{r}} |h_i|$ and the corresponding average SEP is 
\begin{equation}
\begin{aligned}
\SEP^\infty = 1 - \Exp{\br{\qfunc{-\sqrt{\frac{\rho}{N_\textup{r}} ||\h||_1^2}}}^2}. 
\label{eqn: SEP_infty_QPSK}
\end{aligned}
\end{equation}
By the Cauchy-Schwarz inequality, we have
$\|\h\|_1 \le \|\h\|_2\|\1_{N_\textup{r}}\|_2$ and
\begin{align}
\sqrt{\frac{\rho}{N_\textup{r}}}\,\|\h\|_1
\le \sqrt{\rho}\,\|\h\|_2, \nonumber
\end{align}
which implies that the SEP in~\eqref{eqn: new_y_sum_E2G_n_infinity} is always larger than or equal to its unquantized counterpart in~\eqref{eqn: SEP_unq}.
In the limit of $n\to\infty$, Proposition~\ref{prop: mrc_ser_n_high_snr} simplifies to
\begin{subequations}\label{eqn: MRC_high_SNR_infty}
\begin{align}
G_{\dg, \mrc}^\infty &= N_\textup{r}, \label{eqn: MRC_diversity_gain_infty}\\
G_{\cg, \mrc}^\infty &= \frac{(N_\textup{r}!)^{\frac1{N_\textup{r}}}}{N_\textup{r}}. \label{eqn: MRC_high_SNR_coeffi_infty}
\end{align}   
\end{subequations}
\black{Interestingly, from \eqref{eqn: MRC_high_SNR_coeffi_n} and \eqref{eqn: MRC_high_SNR_coeffi_infty}, we observe that the ratio ${G_{\cg, \mrc}^\nbit}/{G_{\cg, \mrc}^\infty} = \frac{\pi}{2^{n-1}}\cot\frac\pi{2^{n-1}}$ rapidly approaches $1$ when $n$ increases. Hence, even with only $n = 4$ bits, the coding gain in~\eqref{eqn: MRC_high_SNR_coeffi_n}
is already about $94.8\%$ of the limit in~\eqref{eqn: MRC_high_SNR_coeffi_infty}.}
\end{remark}

\noindent Fig.~\ref{fig: mrc_n} illustrates the SEP simulation results for a $3$-bit
phase-quantized SIMO-MRC system with $N_\textup{r}\in\{1, 2, 3, 4\}$, corroborating the high-SNR behavior characterized by~\eqref{eqn: MRC_high_SNR_n}. Fig.~\ref{fig: mrc-infty} shows the corresponding results for the limiting case of $n\to\infty$, confirming~\eqref{eqn: SEP_infty_QPSK} and~\eqref{eqn: MRC_high_SNR_infty}.

\begin{figure}[!t]
\centering
\begin{tikzpicture} [auto]
\begin{semilogyaxis}
[
width=8cm,
height=6cm,
ymax=5*10^(-1),
ymin=5*10^(-7),
xmin=0,
xmax=20,
xlabel = {$\rho$ [dB]},
ylabel = {Average SEP},
ylabel near ticks,
x label style={font=\scriptsize},
y label style={font=\scriptsize},
ticklabel style={font=\scriptsize},
legend style = {font=\scriptsize, fill=white, fill opacity=0.6, text opacity=1},
legend pos = south west,
grid=both,
major grid style={line width=.2pt,draw=gray!30},
mark options = {solid},
]

\addplot    [thin, dotted, mark=x, mark repeat=1] table [ y=ser, x=rho, col sep=comma ] {data/simoNr1M4n3.txt}; \addlegendentry{Simulated}
\addplot    [thin, dashed, mark=o, mark repeat=1] table [ y=exact, x=rho, col sep=comma ] {data/simoNr1M4n3.txt}; \addlegendentry{Theoretical \eqref{eqn: mrc_ser_n}}
\addplot    [thin, solid]  table [ y=bound, x=rho, col sep=comma ] {data/simoNr1M4n3.txt}; \addlegendentry{Bound \eqref{eqn: MRC_high_SNR_n}}

\addplot    [thin, dotted, mark=x, mark repeat=1] table [ y=ser, x=rho, col sep=comma ] {data/simoNr2M4n3.txt}; 
\addplot    [thin, dashed, mark=o, mark repeat=1] table [ y=exact, x=rho, col sep=comma ] {data/simoNr2M4n3.txt}; 
\addplot    [thin, solid]  table [ y=bound, x=rho, col sep=comma ] {data/simoNr2M4n3.txt}; 

\addplot    [thin, dotted, mark=x, mark repeat=1] table [ y=ser, x=rho, col sep=comma ] {data/simoNr3M4n3.txt}; 
\addplot    [thin, dashed, mark=o, mark repeat=1] table [ y=exact, x=rho, col sep=comma ] {data/simoNr3M4n3.txt}; 
\addplot    [thin, solid]  table [ y=bound, x=rho, col sep=comma ] {data/simoNr3M4n3.txt}; 

\addplot    [thin, dotted, mark=x, mark repeat=1] table [ y=ser, x=rho, col sep=comma ] {data/simoNr4M4n3.txt}; 
\addplot    [thin, dashed, mark=o, mark repeat=1] table [ y=exact, x=rho, col sep=comma ] {data/simoNr4M4n3.txt}; 
\addplot    [thin, solid]  table [ y=bound, x=rho, col sep=comma ] {data/simoNr4M4n3.txt};

\node[font=\footnotesize, anchor=north west] at (16, 7*10^-2) {$N_\textup{r}=1$};
\node[font=\footnotesize, anchor=north west] at (16, 3*10^-3) {$N_\textup{r}=2$};
\node[font=\footnotesize, anchor=north west] at (16, 1*10^-4) {$N_\textup{r}=3$};
\node[font=\footnotesize, anchor=north west] at (16, 6*10^-6) {$N_\textup{r}=4$};

\end{semilogyaxis}
\end{tikzpicture}
\caption{Average SEP versus $\rho$ for
a  QPSK-modulated, $3$-bit phase-quantized SIMO-MRC system, with $N_\textup{r}\in\{1, 2, 3, 4\}$. The corresponding SEP bound is
specified by~\eqref{eqn:
MRC_high_SNR_n}.}
\label{fig: mrc_n}
\end{figure}

\begin{figure}[!t]
\centering
\begin{tikzpicture} [auto]
\begin{semilogyaxis}
[
width=8cm,
height=6cm,
xmin=0, xmax=20,
ymin=5*10^-9, ymax=1,
xlabel = {$\rho$ [dB]},
ylabel = {Average SEP},
ylabel near ticks,
x label style={font=\scriptsize},
y label style={font=\scriptsize},
ticklabel style={font=\scriptsize},
legend style = {font=\scriptsize},
legend pos = south west,
grid=both,
major grid style={line width=.2pt,draw=gray!30},
mark options = {solid},
]

\addplot    [thin, dotted, mark=x, mark repeat=1] table [ y=ser-sim,   x=rho, col sep=comma ] {data/simoNr1M4Infty.txt} ;\addlegendentry{Simulated}
\addplot    [thin, dashed, mark=o, mark repeat=1] table [ y=ser,   x=rho, col sep=comma ] {data/simoNr1M4Infty.txt} ;\addlegendentry{Theoretical \eqref{eqn: SEP_infty_QPSK}}
\addplot    [thin, solid]  table [ y expr=1^1/factorial(1)*10^(-1*\thisrow{rho}/10), x=rho, col sep=comma ] {data/simoNr1M4Infty.txt};\addlegendentry{Bound \eqref{eqn: MRC_high_SNR_infty}}

\addplot    [thin, dotted, mark=x, mark repeat=1] table [ y=ser-sim,   x=rho, col sep=comma ] {data/simoNr2M4Infty.txt} ;
\addplot    [thin, dashed, mark=o, mark repeat=1] table [ y=ser,   x=rho, col sep=comma ] {data/simoNr2M4Infty.txt} ;
\addplot    [thin, solid]  table [ y expr=2^2/factorial(2)*10^(-2*\thisrow{rho}/10), x=rho, col sep=comma ] {data/simoNr2M4Infty.txt};

\addplot    [thin, dotted, mark=x, mark repeat=1] table [ y=ser-sim,   x=rho, col sep=comma ] {data/simoNr4M4Infty.txt} ;
\addplot    [thin, dashed, mark=o, mark repeat=1] table [ y=ser,   x=rho, col sep=comma ] {data/simoNr4M4Infty.txt} ;
\addplot    [thin, solid]  table [ y expr=4^4/factorial(4)*10^(-4*\thisrow{rho}/10), x=rho, col sep=comma ] {data/simoNr4M4Infty.txt};

\addplot    [thin, dotted, mark=x, mark repeat=1] table [ y=ser-sim,   x=rho, col sep=comma ] {data/simoNr8M4Infty.txt} ;
\addplot    [thin, dashed, mark=o, mark repeat=1] table [ y=ser,   x=rho, col sep=comma ] {data/simoNr8M4Infty.txt} ;
\addplot    [thin, solid]  table [ y expr=8^8/factorial(8)*10^(-8*\thisrow{rho}/10), x=rho, col sep=comma ] {data/simoNr8M4Infty.txt};

\node[font=\footnotesize, anchor=north west] at (14, 3*10^-1) {$N_\textup{r}=1$};
\node[font=\footnotesize, anchor=north west] at (14, 2*10^-2) {$N_\textup{r}=2$};
\node[font=\footnotesize, anchor=north west] at (14, 8*10^-5) {$N_\textup{r}=4$};
\node[font=\footnotesize, anchor=north west] at (14, 5*10^-8) {$N_\textup{r}=8$};

\end{semilogyaxis}
\end{tikzpicture}
\caption{Average SEP versus $\rho$ for a  QPSK-modulated, infinite-bit
phase-quantized SIMO-MRC system, with $N_\textup{r}\in\{1, 2, 4, 8\}$. The
corresponding SEP bound is specified by~\eqref{eqn: MRC_high_SNR_infty}.}
\label{fig: mrc-infty}
\end{figure}


With QPSK modulation ($m=2$), the diversity gains for $n\in\{1,2\}$ and for $n\ge3$ satisfy
\begin{align}
G_{\dg, \mrc}^\onebit=0,\quad G_{\dg, \mrc}^\twobit=\tfrac{N_\textup{r}}{2},\quad G_{\dg, \mrc}^\nbit=N_\textup{r},\nonumber
\end{align}
revealing two distinct phase transitions  around $n=m=2$.
Similar phenomena have been reported in~\cite{Gayan_2020_Open_J, Gay21,
Wu_Liu_et_al_2023} for phase-quantized SIMO-SC and MISO-MRT
systems. These transitions originate from abrupt changes in the key statistics of the SEP expression in~\eqref{eqn: MRC_theorem_proof_2}. In particular, the pdf of the constituent
random variables $Z_i$'s varies significantly
with $n$, as illustrated in Fig.~\ref{fig: pdfs} and discussed in Appendix~\ref{app: key_stats}. 
Its behavior around $0$ is especially revealing:  
\begin{itemize}
    \item For $n = 1$, the pdf
of $Z_i$ has support over the entire $\real$, including negative values, which leads to an error floor and $G_{\mathrm{d}}=0$. 
\item For $n = 2$, $Z_i$ becomes non-negative but its pdf remains strictly positive at $0$ (see~\eqref{eqn: half_normal} in Appendix~\ref{app: key_stats}). 
\item For $n \ge 3$, the pdf of
$Z_i$ is zero at $0$ and the diversity gain behaves as in the unquantized case.
\end{itemize}

\begin{figure}[!t]
\centering
\begin{tikzpicture} [auto]
\begin{axis}
[
width=8cm,
height=6cm,
ymax=1,
ymin=0,
xmin=-2,
xmax=3,
xlabel = $z$,
ylabel = $f_{Z_i}(z)$,
ylabel near ticks,
x label style={font=\scriptsize},
y label style={font=\scriptsize},
ticklabel style={font=\scriptsize},
legend style = {font=\scriptsize, fill=white, fill opacity=0.6, text opacity=1},
legend pos = north west,
grid=both,
major grid style={line width=.2pt,draw=gray!30},
]

\addplot    [name path= flo, thin, dotted, forget plot] table     [ y expr=0*\thisrow{"x"}, x="x", col sep=comma ] {data/pdfs.csv};

\addplot    [name path= one, thin, dotted]        table [ y=1,          x="x", col sep=comma ] {data/pdfs.csv}; \addlegendentry{$n=1$ \eqref{eqn: pdf_Z_n1}}
\addplot    [name path= two, thin, dashed]        table [ y=2,          x="x", col sep=comma ] {data/pdfs.csv}; \addlegendentry{$n=2$ \eqref{eqn: half_normal}}
\addplot    [name path= thr, thin, solid]         table [ y=3,          x="x", col sep=comma ] {data/pdfs.csv}; \addlegendentry{$n=3$ \eqref{eqn: pdf_Z_nge2}}
\addplot    [name path= inf, thin, dashdotted]    table [ y="Infinity", x="x", col sep=comma ] {data/pdfs.csv}; \addlegendentry{$n\rightarrow\infty$ \eqref{eqn: pdf_Z_ninf}}

\addplot[color=gray,opacity=.2] fill between[of = one and flo];
\addplot[color=gray,opacity=.2] fill between[of = two and flo];
\addplot[color=gray,opacity=.2] fill between[of = thr and flo];
\addplot[color=gray,opacity=.2] fill between[of = inf and flo];
\end{axis}
\end{tikzpicture}
\caption{Evolution of the pdf of $Z_i$ in~\eqref{eqn: Z_def} with $n\in\{1, 2, 3\}$ and $n\to\infty$.}
\label{fig: pdfs}
\end{figure}

For an unquantized system, the key statistic in~\eqref{eqn: SEP_unq} is
$\|\h\|^2_2$, which is a scaled chi-square random variable with $2N_\textup{r}$ degrees of freedom (DoF), 
providing diversity gain of $N_\textup{r}$~\cite{Gia03, Sim00, Wu_Liu_et_al_2023}.
In contrast, for $n=2$, the statistic $U$ in~\eqref{eqn: mrc_ser} satisfies
\begin{align}
\frac1{N_\textup{r}}\sum_i Z_i^2 \le U \le \sum_i Z_i^2,\label{eqn: evo_n2}
\end{align}
where the lower bound follows from $Z_i \ge 0$, for $i=1, \ldots, N_\textup{r}$, and the upper bound is based on Cauchy-Schwarz inequality. 
As per~\cite[Prop.~1]{Gia03} (see Appendix~\ref{app: dc}), the achieved diversity gain in this case should be
that of $\sum_i Z_i^2$. However, for $n=2$, $\sum_i Z_i^2$ is a chi-square 
random variable with $N_\textup{r}$ DoF (see Remark~\ref{rem: chi2} in Appendix~\ref{app: key_stats}), thereby yielding a diversity gain of $\frac{N_\textup{r}}{2}$.
Furthermore, for $n\ge3$, define $\alpha_\lb \triangleq \sqrt{1 - \sin\frac{\pi}{2^{n-1}}}$ and $\alpha_\ub \triangleq \sqrt{1 +
\sin\frac{\pi}{2^{n-1}}}$. Similar to~\eqref{eqn: evo_n2}, for the key statistic in~\eqref{eqn: mrc_ser_n}, we have
\begin{align}
\frac{
\alpha_\lb^2
}{N_\textup{r}} 
\sum_i |h_i|^2 \le U \le \alpha_\ub^2 \sum_i |h_i|^2, \nonumber
\end{align}
where the achievable diversity gain is $N_\textup{r}$ as for $\|\h\|^2_2$.

\begin{remark}
Following an approach similar to the proof of~\eqref{eqn: mrc_ser} based on Theorem~\ref{theorem: equivalence_MRC_AO}, it can be shown that the average SEP for a binary PSK-modulated, $2$-bit phase-quantized SIMO-MRC system is
\begin{align}
\SEPBPSK^\twobit  &= 1- \Exp{\qfunc{-\sqrt{\frac{\rho}{N_\textup{r}}\br{\sum_{i=1}^{N_\textup{r}}(|\re(h_i)| + |\im(h_i)|)}^2}}}\nonumber
 \\               
&= \Exp{\qfunc{\sqrt{\frac{\rho}{N_\textup{r}}\br{\norm{\re(\h)}_1+\norm{\im(\h)}_1}^2}}}.\nonumber
\end{align}
As in~\eqref{eqn: evo_n2}, $\br{\norm{\re(\h)}_1+\norm{\im(\h)}_1}^2$ can be  upper- and lower-bounded by scaled chi-square random
variables with $2N_\textup{r}$ DoF as 
\begin{align}
\|\h\|_2^2
\le \br{\norm{\re(\h)}_1+\norm{\im(\h)}_1}^2
\le 2N_\textup{r}\|\h\|_2^2,\nonumber
\end{align}
thereby yielding a diversity gain of $N_\textup{r}$.
\end{remark}

\subsection{Approximate Closed-Form SEP for a QPSK-Modulated, \texorpdfstring{$2$}{2}-Bit Phase-Quantized SIMO-MRC System} \label{sec: closed_form_approx}

The exact expression for the SEP with QPSK inputs and 1-bit
ADCs in~\eqref{eqn:
mrc_ser} is not available in closed form. However, in practice,
having a closed-form approximation can be advantageous. 
For the
special case with $n = 2$, the random variable $U$ appearing in~\eqref{eqn: U_def} equals the square of the (normalized) sum
of i.i.d. half-normal random variables (see~\eqref{eqn: half_normal} in Appendix~\ref{app: key_stats}). To the best of our knowledge, detailed expressions for the
exact pdf and moment generating function of $U$ in this case
are not known, except for $N_r=1$.

For $N_\textup{r} = 1$, the pdf of $U = Z_1^2$ follows a gamma distribution
with both its shape and rate parameters equal to $\frac12$; see Appendix~\ref{app: key_stats}. For general $N_\textup{r}$, using the
structure $U=\frac1{N_\textup{r}}(\sum_i Z_i)^2$ in~\eqref{eqn: U_def} and the fact that a half-normal
random variable $Z_i=\sqrt2|h_i|\cos(\widetilde\theta_i+\frac\pi4)$ is non-negative for $n=2$ (see~\eqref{eqn: Z_def_nge2} in Appendix~\ref{app: key_stats}), we approximate the pdf of $U$ by
a gamma distribution via the well-known
moment-matching method~\cite{Proakis_Dig_comm}. The gamma distribution is parametrized by the shape $\alpha$ and
rate $\lambda$, chosen such that its first two moments coincide with the exact mean and variance
of $U$.

For $n=2$,  the first four moments of $Z_i$ are given by
\begin{align}
\Exp{Z_i} = \sqrt{\frac2\pi},\ \Exp{Z_i^2} = 1,\nonumber\\
\Exp{Z_i^3} = 2\sqrt{\frac2\pi},\ \Exp{Z_i^4} = 3, \nonumber
\end{align}
respectively. Moreover, the mean and variance of $U$ are given by
\begin{align}
\mu_{U} &= \frac{1}{\pi}(-2+2N_\textup{r}+\pi), \nonumber \\
\sigma_{U}^2 &= \frac{2}{\pi^{2}N_\textup{r}}(4(\pi-3)+4N_\textup{r}^2(\pi-2)+N_\textup{r}(20-8\pi+\pi^2)),\nonumber
\end{align}
respectively.
The moment-matching method requires
\begin{align}
\mu_U = \frac{\alpha}{\lambda}, \quad \sigma_U^2 = \frac{\alpha}{\lambda^2}.\nonumber
\end{align}
By solving the above for $\alpha$ and $\lambda$, 
we obtain the shape and rate parameters of the gamma distribution as
\begin{subequations}\label{eqn: gamma_params}
\begin{align}
    \alpha &= \frac{N_\textup{r} (2 N_\textup{r}+\pi -2)^2}{8 (\pi -2) N_\textup{r}^2+2 \mleft(20-8 \pi +\pi ^2\mright) N_\textup{r}+8 (\pi -3)},\label{eqn: alpha}
\\ \lambda &= \frac{\pi N_\textup{r}  (2 N_\textup{r}+\pi -2)}{8 (\pi -2) N_\textup{r}^2+2 \mleft(20-8 \pi +\pi ^2\mright) N_\textup{r}+8 (\pi -3)},\label{eqn: lambda}
\end{align}
\end{subequations}
respectively.

Let $U_\gamma$ be a gamma random variable with parameters $\alpha$ and $\lambda$ given by~\eqref{eqn: gamma_params}. We write~\eqref{eqn: mrc_ser} in the form
\begin{align}
\SEP^\twobit = 2P - P^2,\nonumber
\end{align}
with
\begin{align}
P \triangleq \Exp{\qfunc{\sqrt{\rho U}}} \begin{cases}
                                  = \Exp{\qfunc{\sqrt{\rho U_\gamma}}},\       &N_\textup{r}=1,\\
                                  \approx \Exp{\qfunc{\sqrt{\rho U_\gamma}}},\ &N_\textup{r}\ge2.
                                  \end{cases}\nonumber
\end{align}
Now, define
\begin{align}
    \widetilde P \triangleq \Exp{\qfunc{\sqrt{\rho U_\gamma}}}\nonumber. 
\end{align}
Following a similar approach used in~\cite{Shi04}, for $N_\textup{r} \ge 1$, we have 
$\widetilde P = \big(\frac{\lambda}{2\rho}\big)^{\alpha}\frac{\Gamma(2\alpha)}{{\Gamma(\alpha)}\Gamma(\alpha+1)}\tfo\big(\alpha, \alpha+\frac12; \alpha+1; -\frac{2\lambda}\rho\big)$.
In particular, for $N_\textup{r} = 1$, $\widetilde P$ simplifies to~\cite{Gayan_2020_Open_J}
\begin{align}
\widetilde P|_{N_\textup{r}=1} = \widetilde P_1 \triangleq \frac1\pi\arctan\bigg(\frac1{\sqrt\rho}\bigg) \label{eqn: n2exp},
\end{align}
which leads to
\begin{align}
\SEP^\twobit\begin{cases}
            =2\widetilde P_1  - \widetilde P_1^2,\ &N_\textup{r}=1,\\
            \approx2\widetilde P  - \widetilde P^2,\ &N_\textup{r}\ge2.
            \end{cases}\label{eqn: mrc_ser_cf_approx}
\end{align}
Fig.~\ref{fig: cf-approx} demonstrates that the approximation
is accurate in the low-to-medium SNR range, while somewhat pessimistic at high SNR. The results confirm
that~\eqref{eqn: mrc_ser_cf_approx} is exact for $N_\textup{r} = 1$ and shows that
approximation becomes more accurate as $N_\textup{r}$ increases. Building on this observation, one can straightforwardly combine~\eqref{eqn: mrc_ser_cf_approx} and~\eqref{eqn: MRC_diversity_gain}--\eqref{eqn: MRC_high_SNR_coeffi} to obtain
a more refined approximation for the entire SNR region, given by
\begin{align}
\SEP^\twobit \approx \min\br{2\widetilde P - \widetilde P^2, \br{G_{\cg, \mrc}^\twobit\rho}^{-G_{\dg, \mrc}^\twobit}}.\nonumber
\end{align}
\begin{figure}[!t]
\centering
\begin{tikzpicture} [auto]
\begin{semilogyaxis}
[
width=8cm,
height=6cm,
ymax=10^0,
ymin=10^(-8),
xmin=0,
xmax=20,
ytick={10^0,10^-2,10^-4,10^-6, 10^-8},
xlabel = {$\rho$ [dB]},
ylabel = {Average SEP},
ylabel near ticks,
x label style={font=\scriptsize},
y label style={font=\scriptsize},
ticklabel style={font=\scriptsize},
legend style = {font=\scriptsize, fill=white, fill opacity=0.6, text opacity=1},
legend pos = south west,
grid=both,
major grid style={line width=.2pt,draw=gray!30},
mark options = {solid},
]

\addplot    [thin, dashed, mark=x, mark repeat=2] table [ y=exact, x=rho, col sep=comma ] {data/mrcNr1.txt}; \addlegendentry{Theoretical \eqref{eqn: mrc_ser}}
\addplot    [thin, dotted, mark=triangle, mark repeat=2]  table [ y=ser, x=rho, col sep=comma ] {data/cf_approx_Nr1.txt}; \addlegendentry{Approx. \eqref{eqn: mrc_ser_cf_approx}}
\addplot    [thin, solid]  table [ y=bound, x=rho, col sep=comma ] {data/mrcNr1.txt}; \addlegendentry{Bound \eqref{eqn: MRC_high_SNR}}


\addplot    [thin, dashed, mark=x, mark repeat=2] table [ y=exact, x=rho, col sep=comma ] {data/mrcNr4.txt};
\addplot    [thin, dotted, mark=triangle, mark repeat=2]  table [ y=ser, x=rho, col sep=comma ] {data/cf_approx_Nr4.txt};
\addplot    [thin, solid]  table [ y=bound, x=rho, col sep=comma ] {data/mrcNr4.txt};

\addplot    [thin, dashed, mark=x, mark repeat=2] table [ y=exact, x=rho, col sep=comma ] {data/mrcNr8.txt};
\addplot    [thin, dotted, mark=triangle, mark repeat=2]  table [ y=ser, x=rho, col sep=comma ] {data/cf_approx_Nr8.txt};
\addplot    [thin, solid]  table [ y=bound, x=rho, col sep=comma ] {data/mrcNr8.txt};

\addplot    [thin, dashed, mark=x, mark repeat=2] table [ y=exact, x=rho, col sep=comma ] {data/mrcNr16.txt};
\addplot    [thin, dotted, mark=triangle, mark repeat=2]  table [ y=ser, x=rho, col sep=comma ] {data/cf_approx_Nr16.txt};
\addplot    [thin, solid, domain=0:12] {11.2263959688588 * 10^(-8*x/10)}; 

\node[font=\footnotesize, anchor=north west] at (10.5, 9*10^-1) {$N_\textup{r}=1$};
\node[font=\footnotesize, anchor=north west] at (10.5, 2.5*10^-2) {$N_\textup{r}=4$};
\node[font=\footnotesize, anchor=north west] at (10.5, 3.5*10^-4) {$N_\textup{r}=8$};
\node[font=\footnotesize, anchor=north west] at (10.5, 1*10^-7) {$N_\textup{r}=16$};

\end{semilogyaxis}
\end{tikzpicture}
\caption{Average SEP versus $\rho$ using MRC with $N_\textup{r}\in\{1, 4, 8, 16\}$. The
corresponding SEP bound is specified by~\eqref{eqn: MRC_high_SNR} and the approximate
closed-form SEP specified by~\eqref{eqn: mrc_ser_cf_approx}.}
\label{fig: cf-approx}
\end{figure}

\subsection{Average SEP  for a QPSK-Modulated, \texorpdfstring{$2$}{2}-Bit Phase-Quantized SIMO-SC System}\label{sec: sc}

The diversity gain of a QPSK-modulated, $2$-bit phase-quantized SIMO system under selection combining was established in~\cite[Thm.~4]{Gay21} to be $\frac{N_\textup{r}}{2}$
using a scheme
referred to as the maximum-distance selection, but covering only the cases with $N_\textup{r}= 1, 2$. To fill this gap, we take a fresh perspective on the selection
criterion in~\cite[Sec.~IV-C]{Gay21} and derive the corresponding
diversity and coding gains for arbitrary $N_\textup{r}$.

Note that~\eqref{eqn: mrc_ser} reduces to the SEP of a SISO system for
$N_\textup{r}=1$, which is equivalently expressed as 
\begin{subequations}
\begin{align}
   \SEPSISO &= \Exp{\qfunc{\sqrt{\rho \min(Z_1^2, \widetilde Z_1^2)}}}\nonumber
\\&\phantom= \ +\Exp{\qfunc{\sqrt{\rho\max(Z_1^2, \widetilde Z_1^2)}}} \nonumber
\\&\phantom= \ -\br{\Exp{\qfunc{\sqrt{\rho Z_1^2}}}}^2\label{eqn: SEP_M_eq_1}
        \\&\to \frac{2}{\pi}\rho^{-\frac12}, \ \rho \to \infty, \label{eqn: asymptotic_SISO}   
\end{align}
\end{subequations}
where~\eqref{eqn: asymptotic_SISO} is obtained from~\eqref{eqn:
MRC_diversity_gain}--\eqref{eqn: MRC_high_SNR_coeffi} with $N_\textup{r}=1$. A form equivalent to~\eqref{eqn: SEP_M_eq_1}--\eqref{eqn: asymptotic_SISO}  was also obtained
in~\cite{Gay17}. Furthermore, $\min(Z_1^2, \widetilde Z_1^2)$ can be expressed as
$|h_1|^2(1-|\sin2\widetildetheta_1|)$ (see~\eqref{eqn: Z_def_nge2} in Appendix~\ref{app: key_stats}), whose pdf
is given by (see~\eqref{eq:T_distr} in Appendix~\ref{app: sel_ser})
\begin{align}
f_2(v) \triangleq 2\sqrt{\frac{2}{\pi v}}e^{-\frac{v}{2}}Q[\sqrt{v}], \ v>0. \label{eqn: pdf_2}
\end{align} 
Applying~\cite[Prop.~1]{Gia03} (see Appendix~\ref{app: dc}) to~\eqref{eqn: pdf_2}, we obtain
\begin{align}
\Exp{\qfunc{\sqrt{\rho|h_1|^2(1-|\sin2\widetildetheta_1|))}}} \to \frac{2}{\pi}\rho^{-\frac12}, \ \rho\to\infty.  \label{eqn: discussions_SC}
\end{align}
Comparing~\eqref{eqn: asymptotic_SISO} and~\eqref{eqn: discussions_SC}, we
conclude that the first term on the RHS of~\eqref{eqn: SEP_M_eq_1} dominates
$\SEPSISO$ at high SNR, i.e.,
\begin{align}
\SEPSISO \to \Exp{\qfunc{\sqrt{\rho|h_1|^2(1-|\sin2\widetildetheta_1)|)}}}, \ \rho\to\infty. \label{eqn: discussion_SC_1}
\end{align}     
 
Based on~\eqref{eqn: discussion_SC_1}, to minimize the SEP at relatively high
SNR, we select the index of the antenna branch as
\begin{align}
\argmax_{i\in[N_\textup{r}]}{|h_i|^2(1 - |\sin2\widetildetheta_i|)},\label{eqn: sel_criterion}
\end{align} 
which is equivalent to using the maximum-distance selection in~\cite[Eq.~(13)]{Gay21} for a QPSK-modulated, $2$-bit phase-quantized SIMO-SC system. However, the simple form in~\eqref{eqn: sel_criterion}  would
facilitate further analysis.
Once the antenna branch is selected according to~\eqref{eqn: sel_criterion}, the
detected symbol is obtained solely using the selected antenna branch. The
corresponding SEP takes the same form as~\eqref{eqn: SEP_M_eq_1} except that $h_1$ therein
is replaced by the selected channel coefficient. 

From~\eqref{eqn: discussion_SC_1}--\eqref{eqn: sel_criterion}, the asymptotic average SEP using SC is given by
\begin{align}
\SEP^\sc \to \Exp{\qfunc{\sqrt{\rho\underset{i\in[N_\textup{r}]}{\operatorname{max}}|h_i|^2(1-|\sin2\widetildetheta_i|)}}}, \ \rho\to\infty. \label{eqn: discussion_SC_2}
\end{align}
Using~\eqref{eqn: discussion_SC_2}, we derive the corresponding diversity gain $G_{\dg, \sc}^\twobit$
and coding gain  $G_{\cg, \sc}^\twobit$ in the following proposition.

\begin{proposition}\label{prop: sel_ser}
The diversity and coding gains of a QPSK-modulated, $2$-bit phase-quantized
SIMO-SC system with the selection criteria~\eqref{eqn: sel_criterion} under i.i.d. Rayleigh
fading are given by
\begin{subequations}\label{eqn: Sel_high_SNR}
\begin{align}
G_{\dg, \sc}^\twobit &= \frac{{N_\textup{r}}}{2}, \label{eqn: Sel_diversity_gain}\\
G_{\cg, \sc}^\twobit &= \mleft(2^{2N_\textup{r}-1}\pi^{-\frac{N_\textup{r}+1}{2}}\gammaf{\frac{{N_\textup{r}}+1}{2}}\mright)^{-\frac{2}{N_\textup{r}}},\label{eqn: Sel_high_SNR_coeffi}
\end{align}
\end{subequations}
respectively.
\end{proposition}
\begin{IEEEproof}
See Appendix~\ref{app: sel_ser}.
\end{IEEEproof}

\noindent Fig.~\ref{fig: sc} provides the SEP simulation results of a $2$-bit phase-quantized SIMO-SC
system, with $N_\textup{r}\in\{1, 2, 4, 8\}$, which corroborate~\eqref{eqn: discussion_SC_2}--\eqref{eqn: Sel_high_SNR}.
\begin{figure}[!t]
\centering
\begin{tikzpicture} [auto]
\begin{semilogyaxis}
[
width=8cm,
height=6cm,
xmin=0, xmax=30,
ymin=10^-8, ymax=10^0,
 ytick={10^0,10^-2,10^-4,10^-6,10^-8},
xlabel = {$\rho$ [dB]},
ylabel = {Average SEP},
ylabel near ticks,
x label style={font=\scriptsize},
y label style={font=\scriptsize},
ticklabel style={font=\scriptsize},
legend style = {font=\scriptsize},
legend pos = south west,
grid=both,
major grid style={line width=.2pt,draw=gray!30},
mark options = {solid},
]

\addplot    [thin, dotted, mark=x, mark repeat=2] table [ y=ser, x=rho, col sep=comma ] {data/selNr1.txt}; \addlegendentry{{\color{kalu}Simulated}}
\addplot    [thin, dashed, mark=o, mark repeat=2] table [ y=serapprox, x=rho, col sep=comma ] {data/selNr1.txt}; \addlegendentry{{\color{kalu}RHS of \eqref{eqn: discussion_SC_2}}}
\addplot    [thin, solid] table [ y=bound, x=rho, col sep=comma ] {data/selNr1.txt}; \addlegendentry{Bound \eqref{eqn: Sel_high_SNR}}

\addplot    [thin, dotted, mark=x, mark repeat=2] table [ y=ser, x=rho, col sep=comma ] {data/selNr2.txt};
\addplot    [thin, dashed, mark=o, mark repeat=2] table [ y=serapprox, x=rho, col sep=comma ] {data/selNr2.txt};
\addplot    [thin, solid] table [ y=bound, x=rho, col sep=comma ] {data/selNr2.txt};

\addplot    [thin, dotted, mark=x, mark repeat=2] table [ y=ser, x=rho, col sep=comma ] {data/selNr4.txt};
\addplot    [thin, dashed, mark=o, mark repeat=2] table [ y=serapprox, x=rho, col sep=comma ] {data/selNr4.txt};
\addplot    [thin, solid] table [ y=bound, x=rho, col sep=comma ] {data/selNr4.txt};

\addplot    [thin, dotted, mark=x, mark repeat=1] table [ y=ser, x=rho, col sep=comma ] {data/selNr8.txt};
\addplot    [thin, dashed, mark=o, mark repeat=1] table [ y=serapprox, x=rho, col sep=comma ] {data/selNr8.txt};
\addplot    [thin, solid] table [ y=bound, x=rho, col sep=comma ] {data/selNr8.txt};

\node[font=\footnotesize, anchor=north west] at (24, 1.6*10^-1) {$N_\textup{r}=1$};
\node[font=\footnotesize, anchor=north west] at (24, 1.6*10^-2) {$N_\textup{r}=2$};
\node[font=\footnotesize, anchor=north west] at (24, 3.4*10^-4) {$N_\textup{r}=4$};
\node[font=\footnotesize, anchor=north west] at (24, 8*10^-7)   {$N_\textup{r}=8$};

\end{semilogyaxis}
\end{tikzpicture}
\caption{Average SEP versus $\rho$ using SC with $N_\textup{r}\in\{1, 2, 4, 8\}$.  The
corresponding SEP bound is specified by~\eqref{eqn: Sel_high_SNR}.}
\label{fig: sc}
\end{figure}
 As in the unquantized case, MRC and SC yield the same diversity gain but
different coding gains for $N_\textup{r}>1$, and have identical SEP performance for $N_r=1$.   

 \begin{remark}\label{remark: mrc-sel-rat}
\black{
For the same diversity gain $N_\textup{r}$,  the coding gain of a QPSK-modulated, $2$-bit phase-quantized SIMO-MRC system with $N_\textup{r}$ receive antennas and that  of its SC counterpart with $N_\textup{r}$ receive antennas are related as
\begin{equation}
\begin{aligned}
\frac{G_{\dg, \mrc}^\twobit(N_\textup{r}) }{G_{\dg, \sc}^\twobit(N_\textup{r}) } &= \frac{2^{2-\frac2{N_{\textup{r}}}}}{N_\textup{r}}\mleft({N_\textup{r}}!\mright)^{\frac2{N_\textup{r}}}. \label{eqn: MRC_Sel_rat}
\end{aligned}
\end{equation}}
 \end{remark}
Fig.~\ref{fig: mrc-sel-rat}  presents the ratio of coding gains in dB of
the two diversity methods  by plotting \eqref{eqn: MRC_Sel_rat} in dB. For
example, when $N_\textup{r}=15$, there is a gain of approximately  $10$ dB  in
transmit SNR from using MRC instead of SC. The advantage of MRC over SC is
shown to increase with $N_\textup{r}$, as expected.
 
\begin{figure}
\centering
\begin{tikzpicture}[auto]
\begin{axis}[
width=8cm,
height=6cm,
xmin=1, xmax=32,
ymin=0, ymax=15,
xtick={1,4,8,...,32},
ylabel = {Ratio of coding gains [dB]},
xlabel = {$N_\textup{r}$},
ylabel near ticks,
x label style={font=\scriptsize},
y label style={font=\scriptsize},
ticklabel style={font=\scriptsize},
grid=both,
major grid style={line width=.2pt,draw=gray!30},
]
\addplot [
    domain=1:32, 
    samples=32, 
    thin,
]
{10*log10(4/x * (factorial(x)/2)^(2/x))};
\end{axis}
\end{tikzpicture}
\caption{Ratio of coding gains of using MRC or SC in \eqref{eqn: MRC_Sel_rat} versus number of receive antennas $N_\textup{r}$~\cite{C}.}
\label{fig: mrc-sel-rat}
\end{figure}

\section{Limited CSIR for QPSK Signal Detection} \label{sec: qcsi}

Up to now, we have assumed perfect CSIR  as in~\cite{Gay21, Wu_Liu_et_al_2023}.
However, with coarse quantization, acquiring full CSIR with high precision is often impractical~\cite{Atzeni_2022}. 
In this section, we investigate the effect of limited CSIR for a QPSK-modulated, $2$-bit phase-quantized SIMO system ($m=2=n$). 

\subsection{Description of the Limited CSIR}\label{sec: icsi-desc}
Recall that, for $N_\textup{r} = 1$, we have \[\widehat{s}_\mrc = \qnbitp{2}{h_1^*r_1} = \qnbitp{2}{|h_1|e^{-j\theta_1}r_1}=
\qnbitp{2}{e^{-j\theta_1}r_1},\] 
with  $\theta_1\in(-\pi, \pi)$ (\emph{cf}. \eqref{eqn: Rayleigh_fading_basic}). Now divide the  support of $\theta_1$ into  four intervals: $\mleft(-\frac{\pi}{4},\frac{\pi}{4}\mright)$, $\mleft(\frac{\pi}{4},\frac{3\pi}{4}\mright)$, $\mleft(-\frac{3\pi}{4}, -\frac{\pi}{4}\mright)$ and $\mleft(\frac{3\pi}{4},\pi\mright)\cup\mleft(-\pi,-\frac{3\pi}{4}\mright)$.
It is not hard to observe that, for a given $r_1$, all the realizations of $\theta_1$ falling into one of the intervals above
produce the same detected symbol $\widehat s_\mrc$.
Thus, without full knowledge of $h_1$ or the precise information on $\theta_1$, the $2$-bit information indicating which of the above four intervals 
contains $\theta_1$ produces the same detector output as with full CSIR. 
Alternatively, define $h_1^\icsi=e^{j\frac{\pi}{4}}\br{\qnbitp{2}{h_1e^{j\frac{\pi}{4}}}}^*$,
which leads to
$h_1^\icsi =e^{jl_1\frac\pi2}$, for $l_1= 0, 1, 2, 3$. Moreover, we have
$\widehat{s}_\mrc = \qnbitp{2}{h_1^*r_1} = \qnbitp{2}{h_1^\icsi r_1}$. Therefore, $h_1^\icsi$  represents the $2$-bit channel information required for co-phasing that yields the same detection performance as full CSIR  in SISO channels.

\black{Motivated by the above, we express} the limited knowledge of $\h$ in the form of $\h^\icsi$, the $i$-th element of which is given by $h_i^\icsi= e^{j\frac{\pi}{4}}\br{\qnbitp{2}{h_ie^{j\frac{\pi}{4}}}}^*=e^{jl_i\frac\pi2}$, for $l_i= 0, 1, 2, 3$ and $i=1, \ldots, N_r$. Since $\h^\icsi$ is a form of phase-quantized $\h$, we also refer to $\h^\icsi$ as the $2$-bit phase-quantized CSIR.



\subsection{Detection with Limited CSIR}

Given the above quantized CSIR, we propose to apply the majority-decision rule~\cite{Van01, Proakis_Dig_comm}. Specifically, for each antenna branch $i$, we form $\hat{s}^\icsi_i \triangleq h_i^\icsi r_i =r_ie^{jl_i\frac\pi2}$, for $i=1, \ldots, N_\textup{r}$, and then take the majority decision separately for the real and imaginary parts as
 \[\hat{s}^\icsi_\md \triangleq \qnbitp{2}{{\sum_i \hat{s}^\icsi_i}} = \qnbitp{2}{{(\h^\icsi)\Trans \r}},\] where a possible tie is broken at random\cite{Proakis_Dig_comm}. 

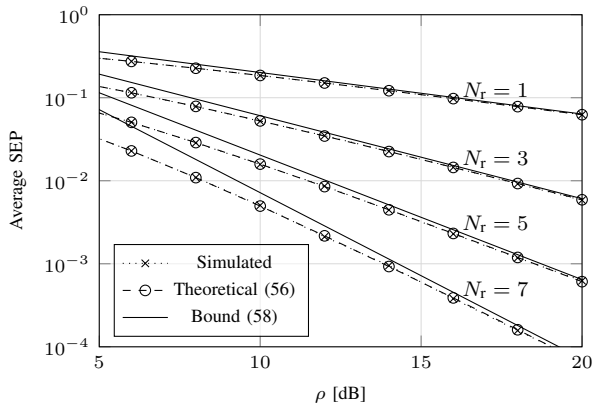
\begin{figure}[!t]
\centering
\begin{tikzpicture} [auto]
\begin{semilogyaxis}
[
width=8cm,
height=6cm,
xmin=0, xmax=20,
ymin=10^-4, ymax=10^0,
xtick={0,5,10,15,20},
xlabel = {$\rho$ [dB]},
ylabel = {Average SEP},
ylabel near ticks,
x label style={font=\scriptsize},
y label style={font=\scriptsize},
ticklabel style={font=\scriptsize},
minor tick style={draw=none},
legend style = {font=\scriptsize, fill=white, fill opacity=0.6, text opacity=1},
legend pos = south west,
grid=major,
major grid style={line width=.2pt,draw=gray!30},
mark options = {solid},
]

\addplot    [thin, dotted, mark=x, mark repeat=1] table [y=ser, x=rho, col sep=comma] {data/icsiNr1.txt};\addlegendentry{Simulated}
\addplot    [thin, dashed, mark=o, mark repeat=1] table [ y=cf, x=rho, col sep=comma ] {data/icsiNr1.txt};\addlegendentry{Theoretical \eqref{eqn: mrc_qcsi_ser}}
\addplot    [thin, solid] table [ y expr=2/pi*(10^(\thisrow{rho}/10))^(-1/2), x=rho, col sep=comma ] {data/icsiNr1.txt};\addlegendentry{Bound \eqref{eqn: QCSI_high_SNR}}

\addplot    [thin, dotted, mark=x, mark repeat=1] table [ y=ser, x=rho, col sep=comma ] {data/icsiNr3.txt};
\addplot    [thin, dashed, mark=o, mark repeat=1] table [ y=cf, x=rho, col sep=comma ] {data/icsiNr3.txt};
\addplot    [thin, solid] table [ y expr=6/pi^2*(10^(\thisrow{rho}/10))^(-2/2), x=rho, col sep=comma ] {data/icsiNr3.txt};

\addplot    [thin, dotted, mark=x, mark repeat=1] table [ y=ser, x=rho, col sep=comma ] {data/icsiNr5.txt};
\addplot    [thin, dashed, mark=o, mark repeat=1] table [ y=cf, x=rho, col sep=comma ] {data/icsiNr5.txt};
\addplot    [thin, solid] table [ y expr=20/pi^3*(10^(\thisrow{rho}/10))^(-3/2), x=rho, col sep=comma ] {data/icsiNr5.txt};

\addplot    [thin, dotted, mark=x, mark repeat=1] table [ y=ser, x=rho, col sep=comma ] {data/icsiNr7.txt};
\addplot    [thin, dashed, mark=o, mark repeat=1] table [ y=cf, x=rho, col sep=comma ] {data/icsiNr7.txt};
\addplot    [thin, solid] table [ y expr=70/pi^4*(10^(\thisrow{rho}/10))^(-4/2), x=rho, col sep=comma ] {data/icsiNr7.txt};

\node[font=\footnotesize, anchor=north west] at (16, 2*10^-1)   {$N_\textup{r}=1$};
\node[font=\footnotesize, anchor=north west] at (16, 3*10^-2)   {$N_\textup{r}=3$};
\node[font=\footnotesize, anchor=north west] at (16, 5*10^-3)   {$N_\textup{r}=5$};
\node[font=\footnotesize, anchor=north west] at (16, 8*10^-4)   {$N_\textup{r}=7$};

\end{semilogyaxis}
\end{tikzpicture}
\caption{SEP with limited CSIR for $N_\textup{r}\in\{1, 3, 5, 7\}$ using the majority-decision rule. The high-SNR characterization follows~\eqref{eqn: QCSI_high_SNR}.}
\label{fig: icsi}
\end{figure}

{\color{kalu}
\begin{proposition}\label{prop: qcsi}
For a QPSK-modulated, $2$-bit phase-quantized SIMO
system with $2$-bit phase-quantized CSIR and using the majority-decision rule, the closed-form average SEP is given by
\begin{align}
\SEP^\icsi &= 2P^\icsi - (P^\icsi)^2\label{eqn: mrc_qcsi_ser},
\end{align}
with \begin{align}P^\icsi = 
\sum_{i=\ceil*{\frac{N_\textup{r}}{2}}}^{2\ceil*{\frac{N_\textup{r}}{2}}-1} 
\ncrr{2\ceil*{\frac{N_\textup{r}}{2}}-1}{i} \widetilde P_1^i(1-\widetilde P_1)^{2\ceil*{\frac{N_\textup{r}}{2}}-1-i},\label{eqn: mrc_qcsi_ser_key_part}
\end{align}
%
%
and where $\widetilde P_1$ is given in~\eqref{eqn: n2exp}.
The corresponding diversity gain $G_{\dg, \md}^\icsi$ and coding gain $G_{\cg, \md}^\icsi$ are given by
\begin{align}
     G_{\dg, \md}^\icsi &= \frac12 \ceil*{\frac{N_\textup{r}}{2}},\quad
     G_{\cg, \md}^\icsi = \pi^2\ncrr{2\ceil*
     {\frac{N_\textup{r}}{2}}}{\ceil*
     {\frac{N_\textup{r}}{2}}}^{-\frac2{\ceil*
     {\frac{N_\textup{r}}{2}}}}, \label{eqn: QCSI_high_SNR}
\end{align}
respectively.
\end{proposition}
\begin{IEEEproof}
See Appendix~\ref{app: icsi-diversity}.
\end{IEEEproof}
}
\noindent Note that~\eqref{eqn: mrc_qcsi_ser}
reduces to~\eqref{eqn: mrc_ser} with $N_\textup{r} = 1$, confirming that co-phasing with $2$-bit phase-quantized CSIR in a QPSK-modulated, $2$-bit phase-quantized SISO system incurs no SEP loss compared with its counterpart with perfect CSIR. 
However, combining multiple antenna branches via the majority-decision rule with limited CSIR clearly
incurs a loss compared with MRC with perfect CSIR. Fig.~\ref{fig: icsi} illustrates the simulation results validating Proposition~\ref{prop: qcsi} along with the corresponding high-SNR bound
from~\eqref{eqn: QCSI_high_SNR}.

\begin{remark}
From~\eqref{eqn: MRC_diversity_gain}, we observe that  $2$-bit phase quantization of
the received signal $\y$ halves the diversity gain compared with its unquantized counterpart. From~\eqref{eqn:
QCSI_high_SNR}, further $2$-bit phase quantization of the channel at the combiner generally incurs  an additional loss of half of  the diversity gain.\end{remark}

\textcolor{black}{Extending Proposition~\ref{prop: qcsi} to a fully general $M$-PSK-modulated, $n$-bit phase-quantized SIMO system with $p$-bit phase-quantized CSIR would require a dedicated and substantially more involved analysis, which we leave for future work.}

\section{Conclusions} \label{sec: concl}

In this paper, we analyzed the SEP for an $M$-PSK-modulated, $n$-bit phase-quantized
SIMO system with $N_\textup{r}$ antennas over i.i.d. Rayleigh fading channels. A key contribution was a
novel approach to the SEP analysis, which leverages the circular symmetry of both the fading and AWGN distributions. Using this approach, we derived exact
analytical SEP expressions for a QPSK-modulated, $n$-bit phase-quantized SIMO-MRC
system, along with the associated diversity and coding gains. Notably, with only $n = 4$ bits, the system achieves full diversity and approximately $94.8\%$
of the coding gain attainable in the unquantized case. In addition to
complementing the existing diversity results on a QPSK-modulated, $2$-bit
phase-quantized SIMO-SC system, we derived an approximate closed-form SEP
expression for the corresponding SIMO-MRC system. 
The proposed analytical framework also provides insights into the MISO-SIMO duality, which we leveraged to characterize the
diversity gain of a general $M$-PSK-modulated, $n$-bit phase-quantized SIMO-MRC
system and to directly transfer these results to the dual MISO setting. Furthermore, we
quantified the additional loss in diversity and coding gains for a QPSK-modulated, $2$-bit
phase-quantized SIMO system incurred when only $2$ bits of
channel phase information per antenna are available at the receiver. All the above
analytical results were verified through simulations. Finally, we note that the SEP characterization of the phase-quantized SIMO-MRC system in Theorem~\ref{theorem: equivalence_MRC_AO} relies on the assumption of i.i.d. channels and noise samples, and may not directly extend to general correlated channels. \black{Computing the coding gain for a phase-quantized SIMO system with correlated channels where $M=2^n$ and a full analysis with limited CSIR remain open problems for future investigation}. 
\appendices

\section{On Diversity and Coding Gains} \label{app: dc}

For the sole purpose of making this paper self-contained, we summarize~\cite[Prop.~1]{Gia03} as follows. To proceed, the following assumptions are required~\cite{Gia03}. 
\begin{enumerate}
\item The instantaneous SNR at the receiver is given by $\rho V$, where $\rho$ is a
positive deterministic constant representing the transmit SNR, and $V$ is a channel-dependent nonnegative random variable.
\item The instantaneous SEP conditioned on $V$ is expressed as $\SEPG(V) = \qfunc{\sqrt{k\rho V}}$, where $k>0$ is a deterministic constant.
\item The pdf of $V$ can be expressed as $f_V(v) = a v^t +
\smallO{v^{t+\epsilon}}$ for $\ v\to0^+$, where $\epsilon>0$ and $a > 0$. The constants $a$ and $t$ are fixed parameters pertaining to the pdf of $V$.
\end{enumerate}
Under the above three assumptions, the asymptotic average SEP at high SNR is characterized as\begin{align}
\SEPG = (G_\cg\rho)^{-G_\dg} + \smallO{\rho^{-G_\dg}},\ \rho\to\infty\nonumber
\end{align}
where $G_\dg \triangleq t+1$ is referred to as the diversity gain, and $G_\cg \triangleq k\mleft(\frac{\smash{2^t}a\gammaf{t+\frac32}}{\smash{\sqrt\pi}(t+1)}\mright)^{-\frac1{t+1}}$ the coding gain.
\black{The derivation steps of above can be found in \cite[Prop.~1]{Gia03}}.

It is clear from the above that the specific values of the deterministic
positive scaling factors $a$ and $k$ do not affect the diversity gain.

\section{Proof of Proposition~\ref{prop: coding_general}}\label{app: coding_general}
{\color{black}
First, we derive the coding gain for a system with the average SEP $\SEPG_\lb=\Exp{\qfunc{\sqrt{\rho}\widetilde T}}$, with $\widetilde T$ defined in \eqref{eqn: MPSK_key_stat}. In the following, we separately consider the cases with $M = 2^n$ and $M < 2^n$.

\subsection{The Case with \texorpdfstring{$M = 2^n$}{M=2 \^ n}}
For $M = 2^n$, using similar derivations as in \cite[Lem.~1]{Wu_Liu_et_al_2023}, it is straightforward to show that the pdf of $\frac{\widetilde Z_i}{\sqrt{2}} = |h_i|\ssin\br{\widetilde \theta_i+\frac \pi M }$ is given by
\begin{align}
f_{\frac{\widetilde Z_i}{\sqrt{2}}}(x) 
= \frac{M}{\sqrt\pi}e^{-x^2}\qfunc{\sqrt 2 x\cot\br{\frac{2\pi}{M}}}.\nonumber
\end{align}
Since our main tool \cite{Gia03} allows us only to know the pdf of the key statistic around $0$, we only require to obtain the pdf of 
 $\sqrt{\frac{N_\textup{r}}{2}}\widetilde T = \sum_i^{{N_\textup{r}}}\frac{\widetilde Z_i}{\sqrt 2}$ around $0$. To that end, we invoke \cite[Prop.~2]{Ribeiro_Cai_Giannakis_2005}
to obtain
\begin{align}
f_{\sqrt{\frac{N_\textup{r}}{2}}\widetilde T}(x) = \frac{1}{({N_\textup{r}}-1)!}\br{\frac{M}{2\sqrt\pi}}^{{N_\textup{r}}}x^{{N_\textup{r}}-1} + \smallO{x^{{N_\textup{r}}-1}},\ x\to0^+.\nonumber
\end{align}
Now, to find the pdf of the key statistic $\frac{N_\textup{r}}{2}\widetilde U = \frac{N_\textup{r}}{2}\widetilde T^2$ around $0$, we have
\begin{align}
f_{\frac{N_\textup{r}}{2}\widetilde U}(x) 
     & = \frac{1}{2\sqrt{x}}f_{\sqrt{\frac{N_\textup{r}}{2}}T}(\sqrt{x})\nonumber
\\ &= \frac{1}{2({N_\textup{r}}-1)!}\br{\frac{M}{2\sqrt\pi}}^{{N_\textup{r}}}x^{\frac{{N_\textup{r}}}{2}-1} + \smallO{x^{\frac{{N_\textup{r}}}{2}-1}},\ x \to 0^+.\nonumber
\end{align}
Finally,~\cite[Prop.~1]{Gia03} is invoked to get 
\begin{align}
    G_{\dg,\mrc,\lb}^\equalcase &= N_\textup{r},\label{eqn: dg_equal_lb}\\
    G_{\cg,\mrc, \lb}^\equalcase &= \br{\frac{M^{N_\textup{r}}2^{-{N_\textup{r}}-1}{N_\textup{r}}^{\frac{N_\textup{r}}2}}{\pi^{\frac{{N_\textup{r}}+1}{2}}{N_\textup{r}}!}\gammaf{\frac{{N_\textup{r}}+1}{2}}}^{-\frac2{N_\textup{r}}}\label{eqn: cg_equal_lb}.
\end{align}

\subsection{The Case with \texorpdfstring{$M < 2^n$}{M<2 \^ n}}
For $M < 2^n$, the distribution of ${\frac{\widetilde Z_i}{\sqrt 2}}$ is 
\begin{align}
\Prob{{\frac{\widetilde Z_i}{\sqrt 2}} \le x}
= 1 - \int_{-\frac\pi {2^n}}^{\frac\pi {2^n}}\frac{2^{n-1}}{\pi}e^{-\frac{x^2}{\sin^2\br{\frac\pi M - \widetilde\theta_i}}}\, d\widetilde\theta_i.\nonumber
\end{align}
By using the dominated convergence theorem \cite[Thm.~(2.27)]{Fol84}, the pdf of ${\frac{\widetilde Z_i}{\sqrt 2}}$ is
\begin{align}
f_{{\frac{\widetilde Z_i}{\sqrt 2}}}(x)
= \frac{2^{n}}{\pi}qx + \smallO{x^3},\ x\to0^+,\nonumber
\end{align}
with $q\triangleq \cot\br{\frac\pi M - \frac\pi{2^n}} - \cot\br{\frac\pi M + \frac\pi{2^n}}$.
By \cite[Thm.~35.1]{Doe63}, the Laplace transform of $f_{{\frac{\widetilde Z_i}{\sqrt 2}}}(x)$ is
\begin{align}
\LT{f_{{\frac{\widetilde Z_i}{\sqrt 2}}}(x)}(s) = \frac{2^{n}}{\pi}q\frac1{s^2} + \smallO{s^{-4}},\ s\to\infty.\nonumber
\end{align}
Since ${\frac{\widetilde Z_i}{2}}$'s are independent, the Laplace transform of the pdf of $\frac{\widetilde T}{\sqrt{2}} = \sum_i {\frac{\widetilde Z_i}{\sqrt 2}}$, $f_{\frac{\widetilde T}{\sqrt 2}}(x)$ is  
\begin{align}
\LT{f_{\frac{\widetilde T}{\sqrt 2}}(x)}(s) &= \br{\LT{f_{{\frac{\widetilde Z_i}{\sqrt 2}}}(x)}(s)}^{N_\textup{r}}\nonumber
\\             &= \br{\frac{2^{n}}{\pi}q}^{N_\textup{r}}\frac1{s^{2{N_\textup{r}}}} + \smallO{s^{-2{N_\textup{r}}-2}},\ s\to\infty.\nonumber
\end{align}
Again applying \cite[Thm.~35.1]{Doe63}, we derive the pdf of $\frac{\widetilde T}{\sqrt{2}}$ as
\begin{align}
f_{\frac{\widetilde T}{\sqrt{2}}}(x) = \br{\frac{2^{n}}{\pi}q}^{N_\textup{r}}\frac1{(2{N_\textup{r}}-1)!}x^{2{N_\textup{r}}-1} + \smallO{x^{2{N_\textup{r}}+1}},\ x\to0^+.\nonumber
\end{align}
Therefore, the pdf of $\frac{N_\textup{r}}{2}U = \frac{N_\textup{r}}{2}T^2$ is
\begin{align}
f_{\frac{N_\textup{r}}{2}U}(x)
= \br{\frac{2^{n}}{\pi}q}^{N_\textup{r}}\frac1{2(2{N_\textup{r}}-1)!}x^{{N_\textup{r}}-1} + \smallO{x^{{N_\textup{r}}}},\ x\to0^+.\nonumber
\end{align}
Now, applying \cite[Prop.~1]{Gia03}, we obtain

\begin{align}
    G_{\dg,\mrc,\lb}^\inequalcase &= N_\textup{r},\label{eqn: dg_inequal_lb}\\
    G_{\cg,\mrc,\lb}^\inequalcase &= \br{\frac{2^{n{N_\textup{r}}-1}N_\textup{r}^{N_\textup{r}}}{{\pi}^{N_\textup{r} + \frac12}(2{N_\textup{r}})!}\gammaf{{N_\textup{r}} + \frac12}\mright.\label{eqn: cg_inequal_lb}\\&\phantom=\mleft.\times\br{\cot\br{\frac\pi M - \frac\pi{2^n}} - \cot\br{\frac\pi M + \frac\pi{2^n}}}^{N_\textup{r}}}^{-\frac1{N_\textup{r}}}.\nonumber
\end{align}    

Up to now, we have obtained the coding gains associated with $\SEPG_\lb$ for the two
cases with $M=2^n$ and with $M<2^n$. To characterize the coding gain of the true $\SEPG$, 
recall the bounds in~\eqref{eqn: MPSK_SEP_bounds}.
Since the diversity gains with $\SEPG_\lb$ and $\SEPG$ are identical, as follows from comparing~\eqref{eqn: general_diversity_gain} with~\eqref{eqn: dg_equal_lb} and~\eqref{eqn: dg_inequal_lb}, we have
\begin{align}
&\br{G_{\cg, \mrc, \lb} \rho}^{-G_{\dg, \mrc}} + \smallO{\rho ^{-G_{\dg, \mrc}}} \nonumber 
\\ \le& \br{G_{\cg, \mrc} \rho}^{-G_{\dg, \mrc}} + \smallO{\rho ^{-G_{\dg, \mrc}}}\nonumber\\\le& 2\br{G_{\cg, \mrc, \lb} \rho}^{-G_{\dg, \mrc}} + \smallO{\rho ^{-G_{\dg, \mrc}}},\ \rho \to \infty\nonumber.
\end{align}
%
Dividing each term by $\rho^{-G_{\dg, \mrc}}>0$ and observing that $\frac{\smallO{\rho^{-G_{\dg, \mrc}}}}{\rho^{-G_{\dg, \mrc}}} \to 0,\ \rho \to \infty$, we obtain
\begin{align}
        &\br{G_{\cg, \mrc, \lb}}^{-G_{\dg, \mrc}} \le \br{G_{\cg, \mrc} }^{-G_{\dg, \mrc}} \le 2\br{G_{\cg, \mrc, \lb}}^{-G_{\dg, \mrc}}\nonumber.
\end{align}
By raising each term of the above to the $-\frac1{G_{\dg, \mrc}}$th power we obtain
\begin{align}
G_{\cg, \mrc, \lb} \ge G_{\cg, \mrc} \ge 2^{-\frac1{G_{\dg, \mrc}}}G_{\cg, \mrc, \lb}\nonumber.  
\end{align}
Hence, there exists a constant $k_{n,M,N_\textup{r}}\in[1,2]$ such that $G_{\cg, \mrc} = k_{n, M, N_\textup{r}}^{-\frac1{G_{\dg, \mrc}}}G_{\cg,\mrc, \lb}$.
Finally, combining the above with \eqref{eqn: cg_equal_lb} and \eqref{eqn: cg_inequal_lb} yields
\eqref{eqn: general_coding_gain}.

}

\section{On the Random Variables Defined in~\texorpdfstring{\eqref{eqn: keystat1}--\eqref{eqn: U_def}}{equations Z def to U def}} \label{app: key_stats}

\begin{remark}\label{rem: ident}
Since we have $\widetilde\theta_i \sim \setU\mleft(-\frac\pi{2^n}, \frac\pi{2^n}\mright)$, it follows directly that $\cos(\widetilde\theta_i + \frac\pi4)$ and $\sin(\widetilde\theta_i + \frac\pi4)$ are identically 
distributed. Consequently, $Z_i$ and $\widetilde Z_i$ have the same distribution, for $i = 1, \ldots, N_\textup{r}$.
As a result, $T$ and $\widetilde T$ are identically distributed, as are $U$ and $\widetilde U$. 
\end{remark}
In what follows, it suffices to focus on the distribution of $Z_i$ when deriving the relevant pdfs.
The following two identities involving the Gaussian $Q$-function are frequently used in our derivations~\cite{Sim00}: for $x\ge0$, we have
\begin{subequations}
\begin{align}
    \qfunc{x} &= \frac1\pi\int_0^{\frac\pi2} e^{-\frac{x^2}{2\sin^2\varphi}}\, d\varphi, \label{eqn: craigs}
\\\qfuncsq{x} &= \frac1\pi\int_0^{\frac\pi4} e^{-\frac{x^2}{2\sin^2\varphi}}\, d\varphi, \label{eqn: Qsq}
\end{align}
\end{subequations}
where~\eqref{eqn: craigs} is commonly known as Craig's formula~\cite{Cra91}. Our subsequent derivations exploit the properties $|h_i|^2 \sim \exp(1)$ and $\widetilde\theta_i \sim \setU\mleft(-\frac\pi{2^n}, \frac\pi{2^n}\mright)$, along with the fact that that $|h_i|$ and $\widetilde\theta_i$
are independent, for $i=1, \ldots, N_r$.
\subsection{The Case with \texorpdfstring{$n=1$}{n=1}}

For $n=1$, we have $\widetilde\theta_i
\sim \setU\mleft(-\frac\pi2, \frac\pi2\mright)$. \black{By utilizing \eqref{eqn: Qsq}}, we obtain
\begin{align}
\Prob{Z_i \le z} &= \Prob{\sqrt2|h_i|\cos\mleft(\widetilde\theta_i + \frac\pi4\mright) \le z}\nonumber
\\               &= \qfuncsq{-z},\ z\in \real.\nonumber
\end{align}
\black{Taking the derivative of the above,} we have that $Z_i$'s are i.i.d. with pdf given by
\begin{align}
f_{Z_i}(z) = \sqrt{\frac{2}{\pi}}\qfunc{-z}e^{-\frac{z^2}2},\ z\in \real.\label{eqn: pdf_Z_n1}
\end{align}

\subsection{The Case with \texorpdfstring{$n\ge 2$}{n >= 2}}

For $n\ge 2$, $Z_i$ and $\widetilde Z_i$ are both nonnegative, with
\begin{align}
Z_i = |h_i|\sqrt{1-\sin2\widetildetheta_i},\quad \widetilde Z_i = |h_i|\sqrt{1+\sin2\widetildetheta_i}. \label{eqn: Z_def_nge2}
\end{align}
Therefore, for $z\ge0$, we have
\begin{align}
\Prob{Z_i \le z} &= \Expp{\widetilde\theta_i}{\Prob{\mleft.|h_i|\sqrt{1 - \sin2\widetildetheta_i}\le z\mright|\widetildetheta_i}}\nonumber
\\               &= 1 - \frac{2^{n-1}}\pi\int_{-\frac\pi{2^n}}^{\frac\pi{2^n}}e^{-\frac{z^2}{1 - \sin2\widetilde\theta_i}}\, d\widetildetheta_i.\label{eqn: cdf_Z_general}
\end{align}
For the corresponding pdf, here we consider the case with $n\ge 3$, and leave the case with $n=2$ to the next subsection. For $n\ge 3$, the required pdf of $Z_i$ can be obtained by taking the derivative of~\eqref{eqn: cdf_Z_general} with respect to $z$.  The derivative may be taken inside the integral by the dominated  convergence theorem~\cite[Thm.~(2.27)]{Fol84}, which gives
\begin{align}
f_{Z_i}(z) &= \frac{2^{n}z}\pi \int_{-\frac\pi{2^n}}^{\frac\pi{2^n}}\frac{e^{-\frac{z^2}{1-\sin2\widetilde\theta_i}}}{1-\sin2\widetilde\theta_i} \, d\widetildetheta_i,\ z\ge0\label{eqn: pdf_Z_nge2}.
\end{align}
Clearly, the above pdf is well defined in its integral form. In the limit of $n\to\infty$, $Z_i$ converges to a Rayleigh random variable with pdf given by
\begin{align}
f_{Z_i}(z) = 2ze^{-z^2},\ z\ge0.\label{eqn: pdf_Z_ninf}
\end{align}

\subsection{The Case with \texorpdfstring{$n=2$}{n=2}}

For $n=2$, \black{using \eqref{eqn: craigs}} and~\eqref{eqn: cdf_Z_general}, we have
\begin{align}
\Prob{Z_i \le z} = 1 - 2\qfunc{z},\ z \ge 0,\nonumber
\end{align}
for $i=1, \ldots, N_\textup{r}$. \black{Taking the derivative of the above}, $Z_i$ follows a half-normal distribution with pdf given by
\begin{align}
f_{Z_i}(z) = \sqrt{\frac{2}{\pi}}e^{-\frac{z^2}2},\ z\ge0.\label{eqn: half_normal}
\end{align}
Previously, we mentioned that $Z_i$ and $\widetilde Z_i$ are identically 
distributed. For $n=2$, we have the following stronger result.

\begin{lemma}\label{lemma: Z_tildeZ_iid}
$Z_i$ and $\widetilde Z_i$ are i.i.d. if and only if $n = 2$ holds.
\end{lemma}
\begin{IEEEproof}
By Remark~\ref{rem: ident}, $Z_i$ and $\widetilde Z_i$ are identically distributed. Now, we focus on proving their independence for $n=2$.
For brevity, define
$X \triangleq Z_i^2$, $Y \triangleq \widetilde Z_i^2$, and $W \triangleq |h_i|^2$.
From ~\eqref{eqn: Z_def_nge2}, we have 
$X = W(1-\sin2\widetilde\theta_i)$
and
$Y = W(1+\sin2\widetilde\theta_i)$.
The characteristic function of the random vector $(X, Y)$ is 
\begin{align}
\varphi_{(X, Y)}(\omega, \nu) \triangleq \Exp{e^{j(\omega X  +\nu Y)}},\nonumber
\end{align}
which is further expanded as
\begin{align}
&\varphi_{(X, Y)}(\omega, \nu) \nonumber
\\&= \Expp{W}{\Expp{\widetilde\theta_i}{e^{j(\omega W(1-\sin2\widetilde\theta_i) + \nu W(1+\sin2\widetilde\theta_i))}}}\nonumber
\\&=\Expp{W}{\Expp{\widetilde\theta_i}{e^{\sin2\widetilde\theta_i\mleft(j(\nu-\omega)W\mright)}}e^{j(\nu+\omega)W}}. \nonumber
\end{align}
Since we have $\widetilde\theta_i\sim\setU\br{-\frac{\pi}{4}, \frac{\pi}{4}}$, it follows that
\begin{align}
    \Expp{\widetilde\theta_i}{e^{\sin2\widetilde\theta_i\mleft(j(\nu-\omega)W\mright)}}&=\frac{2}{\pi}\displaystyle\int_{-\frac{\pi}{4}}^{\frac{\pi}{4}}e^{j(\nu-\omega)W\sin2\widetilde\theta_i}d\widetilde\theta_i  \nonumber
    \\&=\frac{1}{\pi}\displaystyle\int_{0}^{\pi}e^{j(\omega-\nu)W\cos\widetilde\theta_i}d\widetilde\theta_i\nonumber
    \\&=J_0(W(\omega-\nu)). \nonumber
\end{align}
Therefore, we have
\begin{align}
\varphi_{(X, Y)}(\omega, \nu)  
&=\int_0^\infty e^{-(1-j(\omega + \nu))W}J_0(W(\omega - \nu))\, dW\nonumber
%
\\&= \LT{J_0(W(\omega - \nu))}(1-j(\omega + \nu))\nonumber
\\                            &= \frac1{\sqrt{(1-2j\omega)(1-2j\nu)}},\nonumber
\end{align}
where we have used~\cite[Eq.~6.611]{Gra07} to obtain last equality.
On the other hand, from \eqref{eqn: half_normal}, we can obtain the pdfs of $X$ and $Y$, which are identical, as
\begin{align}
\varphi_{X}(\omega) = \frac1{\sqrt{1-2j\omega}},\ \varphi_{Y}(\nu) = \frac1{\sqrt{1-2j\nu}}, \nonumber
\end{align}
i.e., $\varphi_{(X, Y)}(\omega, \nu) = \varphi_X(\omega)\varphi_Y(\nu)$.
Hence, for $n=2$, $X$ and $Y$ are independent~\cite{Papoulis_Pillai_Prob}, and so are $Z_i$ and $\widetilde Z_i$.
For general $n$, the covariance between $Z_i$ and $\widetilde Z_i$ is given by
$ \frac{2^n}\pi\sin\frac\pi{2^n}\br{\cos\frac\pi{2^n} - 2^{n-2}\sin\frac\pi{2^n}}$
which is nonzero for $n\ne2$, implying that $Z_i$ and $\widetilde Z_i$ are dependent for
$n\ne2$. 
\end{IEEEproof}

\begin{remark}
As a result of Lemma~\ref{lemma: Z_tildeZ_iid}, it is clear that
$T$ and $\widetilde T$ as well as $U$ and $\widetilde U$ are independent if and only if $n = 2$ holds.
\end{remark}

\begin{remark}\label{rem: chi2}
The two random variables
$\sqrt2|\re(h_i)|$ and
$\sqrt2|\im(h_i)|$ are i.i.d. with the same pdf as in~\eqref{eqn: half_normal}, for 
$i = 1, \ldots , N_\textup{r}$. Clearly, for $n=2$, the pair of random variables $Z_i$ and $\widetilde Z_i$
can be replaced by
$\sqrt2|\re(h_i)|$ and
$\sqrt2|\im(h_i)|$. In addition, it
is straightforward to observe that 
$Z_i^2$
is a chi-square distributed random
variable with $1$ degree of freedom (i.e., gamma distributed with the shape and rate parameters equal to $\frac12$), whereas $\sum_{i=1}^{N_\textup{r}}Z_i^2$
is chi-square with $N_\textup{r}$ DoF.
\end{remark}


\section{Proof of Proposition~\ref{prop: sel_ser}} \label{app: sel_ser}

Recall~\eqref{eqn: sel_criterion}.  Let $V_i \triangleq |h_i|^2(1 -
|\sin{2\widetildetheta_i}|)$ and $V_{\max} \triangleq
\max_{i}\; V_i$. Then, based on~\eqref{eqn: discussion_SC_2}, we have 
\begin{align}
\SEP^\sc \to \Exp{\qfunc{\sqrt{\rho V_{\max}}}}, \quad \rho\to\infty.\nonumber
\end{align}
For $v\ge0$, we have
\begin{align}
\Prob{V_{\max} \leq v} &= \prod_{i=1}^{N_\textup{r}}\Prob{V_i \leq v} = \br{\Prob{V_i \leq v}}^{N_\textup{r}},\nonumber
\end{align}
which holds due to $V_i$'s being i.i.d. 
Using an approach similar to that of~\eqref{eqn: cdf_Z_general}, we obtain
\begin{align}
\Prob{V_i\leq v} &= \Prob{|h_i|^2(1-|\sin{2\widetildetheta_i|)} < v}\nonumber\\
             &= 1 - \frac4\pi\int_0^{\frac\pi4}e^{-\frac v{2\sin^2\widetilde\theta_i}}\, d\widetildetheta_i\nonumber\\
             &= 1 - 4\br{\qfunc{\sqrt{v}}}^2, \quad v\ge0,\label{eq:T_distr}
\end{align}
where we have utilized~\eqref{eqn: Qsq}.
Thus, the pdf of $V_{\max}$ is given by
\begin{equation}\label{eq:pdf_T_max}
\begin{aligned}
f_{V_{\max}}(v) &= \frac{4{N_\textup{r}}}{\sqrt{2\pi v}} e^{-\frac{v}{2}}\qfunc{\sqrt{v}}\br{1 - 4\br{\qfunc{\sqrt{v}}}^2}^{{N_\textup{r}}-1},\\
\end{aligned}
\end{equation}
for $v\ge0$. 
Next, 
by considering $v\to 0^+$, we have
\begin{equation}\label{eq:Qofsqrtt}
\begin{aligned}
\qfunc{\sqrt v} &= \frac{1}{2} - \frac{1}{\sqrt{2\pi}}v^{\frac{1}{2}} + \smallO{v^{\frac{3}{2}}}.\\
\end{aligned}
\end{equation}
Applying~\eqref{eq:Qofsqrtt} 
to~\eqref{eq:pdf_T_max}, after some algebra, the pdf of $V_{\max}$  for $v \to 0^+$ is given by
\begin{equation}\label{eq:pdf_T_max_approx}
\begin{aligned}
f_{V_{\max}}(v) &= d_1 v^{d_2} + \smallO{v^{d_2}},\\
\end{aligned}
\end{equation}
with
\begin{equation}\label{eqn: Sel_pdf_params}
d_1 \triangleq \frac{N_\textup{r}}2\mleft(\frac{4}{\sqrt{2\pi }}\mright)^{N_\textup{r}}, \quad d_2 \triangleq \frac{N_\textup{r}}{2} -1. 
\end{equation}
Based  on~\eqref{eq:pdf_T_max_approx}--\eqref{eqn: Sel_pdf_params}, using~\cite[Prop.~1]{Gia03}, we can
readily  obtain~\eqref{eqn: Sel_diversity_gain} and~\eqref{eqn:
Sel_high_SNR_coeffi}.

\section{Proof of Proposition~\ref{prop: qcsi}}\label{app: icsi-diversity}

First, we consider the case with odd $N_\textup{r}$. 
According to the  majority-decision rule, if $\ceil*{N_\textup{r}/2}$ of the
antenna branches make the correct individual decisions on the real and imaginary parts of the symbol, then the receiver makes the correct decision.

Let $\text{e}_\re$ (resp. $\text{e}_\im$) denote
the event of an erroneous majority decision on the real
(resp. imaginary) part of the transmitted symbol $s$. 
In this case, 
we have
\begin{align}
\SEP^\icsi     &= 1 - \Exp{\Prob{\text{correct decision}|\h^\icsi}} \nonumber\\
               &= 1 - \Exp{(1 -\Prob{\text{e}_\re|\h^\icsi})(1 -\Prob{\text{e}_\im|\h^\icsi})} \label{eqn: icsi_proof_1}.
\end{align}
\black{In above, \eqref{eqn: icsi_proof_1} follows because $\re(\widehat s_i^\icsi)$ and $\im(\widehat s_i^\icsi)$ are independent when $n=2$}.
As earlier,  we consider $s=e^{j\frac\pi4}$, without loss of generality. Recall~\eqref{eqn: Z_def}.
After some algebra, we obtain
\begin{align}
\Prob{\text{e}_\re|\h^\icsi} &=\!\sum_{\setW\in\ncr{[N_\textup{r}]}{\ge\ceil*{N_\textup{r}/2}}}\! \prod_{i\in \setW} \!p_i           \!\prod_{ k\in [N_\textup{r}]\setminus \setW}\!(1-p_{k})\nonumber,\\
\Prob{\text{e}_\im|\h^\icsi} &=\!\sum_{\setW\in\ncr{[N_\textup{r}]}{\ge\ceil*{N_\textup{r}/2}}}\! \prod_{i\in \setW} \!\widetilde p_i\!\prod_{ k\in [N_\textup{r}]\setminus \setW}\!(1-\widetilde p_{k})\nonumber,
\end{align}
where \black{$\setW$ contains a set of antenna-branch indices that leads to an incorrect decision}, $p_i = \qfunc{\sqrt{\rho}Z_i}$ \black{is the probability of the real part of the decision of the $i$th antenna branch being incorrect} and $\widetilde p_i = \qfunc{\sqrt{\rho}\widetilde Z_i}$ \black{is the probability of the imaginary part of the decision of the $i$th antenna branch being incorrect}.
Since $\Prob{\text{e}_\re|\h^\icsi}$ involves only $Z_i$'s whereas $\Prob{\text{e}_\im|\h^\icsi}$ contains only $\widetilde Z_i$'s, by Lemma~\ref{lemma: Z_tildeZ_iid}, $\Prob{\text{e}_\re|\h^\icsi}$ and $\Prob{\text{e}_\im|\h^\icsi}$ are i.i.d. Correspondingly, by letting
\[P^\icsi =\Exp{\Prob{\text{e}_\re|\h^\icsi}}=\Exp{\Prob{\text{e}_\im|\h^\icsi}},\]
from \eqref{eqn: icsi_proof_1}, we obtain
\begin{align}
\SEP^\icsi = 2\Exp{\Prob{\text{e}_\re|\h^\icsi}} - \br{\Exp{\Prob{\text{e}_\re|\h^\icsi}}}^2,\nonumber
\end{align}
which establishes \eqref{eqn: mrc_qcsi_ser}.

Since  all the channel coefficients 
are i.i.d., after some algebra, we have
\begin{align}
P^\icsi=\Exp{\Prob{\text{e}_\re|\h^\icsi}}
= \sum_{i=\ceil*{N_\textup{r}/2}}^{N_\textup{r}} \ncr{N_\textup{r}}{i} \widetilde P_1^i(1-\widetilde P_1)^{N_\textup{r}-i}\nonumber
\end{align}
with $\widetilde P_1$ given by~\eqref{eqn: n2exp}.  As per the tie-breaking rule, for even $N_r$,  $P^\icsi$ is the same as that for $N_\textup{r} - 1$. Thus,   for all values of $N_r$,  $P^\icsi$ is  
given by \eqref{eqn: mrc_qcsi_ser_key_part}. 

 As $\rho\to\infty$, $\widetilde P_1 \to \frac1\pi\rho^{-\frac12}$. Based on \eqref{eqn: mrc_qcsi_ser}--\eqref{eqn: mrc_qcsi_ser_key_part}, we have
\begin{align}
\SEP^\icsi &= 2\begin{pmatrix}2\ceil*{N_\textup{r}/2}-1\\\ceil*{N_\textup{r}/2}\end{pmatrix} (\pi^2 \rho)^{-\frac12 \ceil*{N_\textup{r}/2}} + \smallO{\rho^{-\frac12 \ceil*{N_\textup{r}/2}}}\nonumber
 \\&= \begin{pmatrix}{2\ceil*{N_\textup{r}/2}}\\{\ceil*{N_\textup{r}/2}}\end{pmatrix} (\pi^2 \rho)^{-\frac12 \ceil*{N_\textup{r}/2}} + \smallO{\rho^{-\frac12 \ceil*{N_\textup{r}/2}}}\nonumber
\end{align}
as $\rho\to\infty$,
from which  the diversity coding gains are clearly given by~\eqref{eqn: QCSI_high_SNR}.

\section{List of Symbols}
Table~\ref{tab:test} lists the symbols used throughout this paper.
\begin{longtblr}[
  caption = {List of symbols.},
  label = {tab:test},
]{
  colspec = {|XX[4]|},
  rowhead = 1,
  hlines,
  row{even} = {gray9},
  row{1} = {olive9},
} 
\textbf{Symbol} & \textbf{Description} \\
$\alpha$ & Shape parameter of gamma distribution \\
$\alpha_\ub$ & Constituent parameter to upper bound $U$ for $M=4$, $n\ge3$ case:$\sqrt{1 + \sin\frac{\pi}{2^{n-1}}}$\\
$\alpha_\lb$ & Constituent parameter to lower bound $U$ for $M=4$, $n\ge3$ case:$\sqrt{1 - \sin\frac{\pi}{2^{n-1}}}$  \\
$\theta_i$ & Channel phase of antenna branch $i$ \\
$\widetilde{\theta}_i$ & Quantization phase error of antenna branch $i$\\
$\lambda$ & Rate parameter of gamma distribution \\
$\mu_U$ & Mean of $U$ \\
$\phi_i$ & Quantized phase of antenna branch $i$ \\
$\rho$ & Transmit SNR \\
$\sigma_U^2$ & Variance of $U$ \\
$\omega$ & Argument of the charcteristic function \\

$\setE_1$ & Error event set of original system \\
$\setE_2$  & Error event set of transformed system \\
$\widetilde{\setE}_2$ & Error event set of SISO interpretation of $\setE_2$\\
$\setE^\miso$ & Error event set of MISO system\\
$G_\cg$ & Coding gain \\
$G_{\cg, \mrc}^\equalcase$ & Coding gain of MRC for $M=2^n$ case\\
$G_{\cg, \mrc}^\inequalcase$ & Coding gain of MRC for $M<2^n$ case\\
$G_{\cg, \mrc, \lb}^\equalcase$ & Coding gain for $\SEPG_\lb$, $M=2^n$ case \\
$G_{\cg, \mrc, \lb}^\inequalcase$ & Coding gain for $\SEPG_\lb$, $M<2^n$ case \\
$G_{\cg, \mrc}^\onebit$ & Coding gain of MRC for $M=4$, $n=1$ case\\
$G_{\cg, \mrc}^\twobit$ & Coding gain of MRC for $M=4$, $n=2$ case\\
$G_{\cg, \mrc}^\nbit$ & Coding gain of MRC for $M=4$, $n\ge2$ case\\
$G_{\cg, \mrc}^\uq$ & Coding gain of MRC for $M=4$, unquantized case\\
$G_{\cg, \mrc}^\infty$ & Coding gain of MRC for $M=4$, infinite-resolution case\\
$G_{\cg, \sc}^\twobit$ & Coding gain of SC for $M=4$, $n=2$ case\\
$G_{\cg, \md}^\icsi$ & Coding gain of MRC for limited-CSI case\\
$G_\dg$ & Diversity gain \\
$G_{\dg, \mrc}$ & Diversity gain for MRC \\
$G_{\dg, \mrc, \lb}^\equalcase$ & Diversity gain for $\SEPG_\lb$, $M=2^n$ case \\
$G_{\dg, \mrc, \lb}^\inequalcase$ & Diversity gain for $\SEPG_\lb$, $M<2^n$ case \\
$G_{\dg, \mrc}^\onebit$ & Diversity gain of MRC for $M=4$, $n=1$ case\\
$G_{\dg, \mrc}^\twobit$ & Diversity gain of MRC for $M=4$, $n=2$ case\\
$G_{\dg, \mrc}^\nbit$ & Diversity gain of MRC for $M=4$, $n\ge2$ case\\
$G_{\dg, \mrc}^\uq$ & Diversity gain of MRC for $M=4$, unquantized case\\
$G_{\dg, \mrc}^\infty$ & Diversity gain of MRC for $M=4a,$ infinite-resolution case\\
$G_{\dg, \sc}^\twobit$ & Diversity gain of SC for $M=4$, $n=2$ case\\
$G_{\dg, \md}^\icsi$ & Diversity gain of MRC for limited-CSI case\\
$\h$ & Channel vector \\
$h_i$ & Channel of antenna branch $i$ \\
$h_i^\icsi$ & Quantized channel phase information of antenna branch $i$ \\
$\widetilde{h}_i$ & Effective channel coefficient of antenna branch $i$\\
$k_{n, M, N_\textup{r}}$ & a real number between $1$ and $2$ \\
$l_i$ & Quantized channel phase sector index of antenna branch $i$ \\
$M$ & Modulation order \\
$m$ & Number of modulation bits, $m=\log_2 M$ \\
$n$ & Quantization resolution in bits \\
$\n$ & Noise vector \\
$n_\summa$ & Equivalent combined noise\\
$\widetilde{n}_i$ & Rotated noise sample \\
$N_\textup{r}$ & Number of receive antennas \\
$N_\textup{t}$ & Number of transmit antennas \\
$\SEPG$ & Average symbol error probability (SEP) \\
$\SEPG^\efcase$ & SEP for $M > 2^n$ case\\
$\SEP^\onebit$ & SEP for $M=4$, $n=1$ case \\
$\SEP^\twobit$ & SEP for $M=4$, $n=2$ case \\
$\SEP^\nbit$ & SEP for $M=4$, $n\ge3$ case \\
$\SEP^\infty$ & SEP for $M=4$, infinite-resolution case \\
$\SEP^\uq$ & SEP for $M=4$, unquantized case \\
$\SEPG_\lb$ & Lower bound on SEP \\
$\SEPG_\miso$ & SEP of the MISO system \\
$\SEPBPSK^\twobit$ & SEP for $M=2$, $n=2$ case \\
$\SEP^\icsi$ & SEP with limited CSIR \\
$\SEP^\sc$ & SEP for SIMO-SC \\
$\SEPSISO$ & SEP for SISO case \\
$P^\icsi$ & Intermediate probability term for limited CSIR \\
$\widetilde P$ & Approximate intermediate SEP term \\
$\widetilde P_1$ & Approximate intermediate SEP term when $N_\textup{r} = 1$ \\
$q$ & Constituent parameter in coding gain derivation\\
$\r$ & Quantized received vector \\
$r_i$ & Quantized received signal of antenna branch $i$ \\
$s$ & Transmitted symbol \\
$\hat{s}_\mrc$ & Detected symbol using MRC \\
$\hat{s}^\icsi_\md$ & Majority-decision detected symbol with limited CSIR \\
$\mathcal{S}_M$ & $M$-PSK constellation \\
$T$ & Decision statistic:$\frac{1}{\sqrt{N_\textup{r}}}\sum_{i=1}^{N_\textup{r}}Z_i$\\
$\widetilde{T}$ & Decision statistic:$\frac{1}{\sqrt{N_\textup{r}}}\sum_{i=1}^{N_\textup{r}}\widetilde Z_i$ \\
$U$ & Squared decision statistic:$T^2$ \\
$\widetilde{U}$ & Squared decision statistic:$\widetilde T^2$ \\
$U_\gamma$ & Gamma-approximated version of $U$ \\
$\widetilde w$ & Noise in MISO system \\
$x$ & Argument in some pdfs \\
$\y$ & Unquantized received signal vector\\
$y_i$ & Unquantized received signal antenna branch $i$\\
$y_\summa$ & Equivalent combined received signal \\
$y_\summa^\infty$ & Equivalent combined received signal in infinite-resolution case\\
$\widetilde{y}_i$ & Rotated received signal of antenna branch $i$:$e^{j\phi_i}y_i$\\
$Z_i$ & Constituent random variable of antenna branch $i$:$\sqrt2|h_i|\cos\br{\frac\pi M + \widetilde \theta_i}$ \\
$\widetilde{Z}_i$ & Constituent random variable of antenna branch $i$:$\sqrt2|h_i|\sin\br{\frac\pi M + \widetilde \theta_i}$ \\
\end{longtblr}

\bibliographystyle{IEEEtran}
\bibliography{IEEEabbr, refsJ}

\end{document}